\documentclass{IEEEtran}
\usepackage{cite}
\usepackage{amsmath,amssymb,amsfonts}
\usepackage{amsthm}
\usepackage{algorithmic}
\usepackage{graphicx}
\graphicspath{{figs_two_column/}{matlab/}{../figs_two_column/}} 
\usepackage[font=footnotesize]{caption}
\usepackage{subcaption}
\usepackage{calc}
\usepackage{multicol}
\usepackage{psfrag}
\usepackage[usenames,dvipsnames]{pstricks}
\usepackage{epsfig}
\usepackage{enumitem}
\usepackage{textcomp}
\def\BibTeX{{\rm B\kern-.05em{\sc i\kern-.025em b}\kern-.08em
    T\kern-.1667em\lower.7ex\hbox{E}\kern-.125emX}}

\newcommand{\cS}{\mathcal{S}}
\newcommand{\cY}{\mathcal{Y}}
\newcommand{\cX}{\mathcal{X}}
\newcommand{\cF}{\mathcal{F}}

\newcommand{\cI}{\mathcal{I}}
\newcommand{\cC}{\mathcal{C}}
\newcommand{\cA}{\mathcal{A}}
\newcommand{\cM}{\mathcal{M}}
\newcommand{\cN}{\mathcal{N}}
\newcommand{\cR}{\mathcal{R}}
\newcommand{\cD}{\mathcal{D}}

\newcommand{\cJ}{\mathcal{J}}
\newcommand{\cT}{\mathcal{T}}
\newcommand{\cB}{\mathcal{B}}
\newcommand{\cH}{\mathcal{H}}
\newcommand{\cP}{\mathcal{P}}
\newcommand{\cQ}{\mathcal{Q}}

\newcommand{\BSC}{\mathrm{BSC}}

\newcommand{\td}[1]{\tilde{#1}}
\newcommand{\abs}[1]{{\left\lvert #1 \right\rvert}}

\theoremstyle{plain}
\newtheorem{theorem}{Theorem}

\newtheorem{lemma}[theorem]{Lemma}

\theoremstyle{definition}
\newtheorem{example}[theorem]{Example}

\newtheorem{definition}[theorem]{Definition}

\usepackage{xspace}
\usepackage{bbm}
\catcode`@=11
\chardef\mathlig@atcode\count255

\def\actively#1#2{\begingroup\uccode`\~=`#2\relax\uppercase{\endgroup#1~}}
\def\mathlig@gobble{\afterassignment\mathlig@next@cmd\let\mathlig@next= }
\def\mathlig@delim{\mathlig@delim}
\def\mathlig@defcs#1{\expandafter\def\csname#1\endcsname}
\def\mathlig@let@cs#1#2{\expandafter\let\expandafter#1\csname#2\endcsname}
\def\mathlig@appendcs#1#2{\expandafter\edef\csname#1\endcsname{\csname#1\endcsname#2}}

\def\mathlig#1#2{\mathlig@checklig#1\mathlig@end\mathlig@defcs{mathlig@back@#1}{#2}\ignorespaces}

\def\mathlig@checklig#1#2\mathlig@end{%
 \expandafter\ifx\csname mathlig@forw@#1\endcsname\relax
 \expandafter\mathchardef\csname mathlig@back@#1\endcsname=\mathcode`#1%
 \mathcode`#1"8000\actively\def#1{\csname mathlig@look@#1\endcsname}%
 \mathlig@dolig#1\mathlig@delim
\fi
\mathlig@checksuffix#1#2\mathlig@end
}

\def\mathlig@checksuffix#1#2\mathlig@end{%
\ifx\mathlig@delim#2\mathlig@delim\relax\else\mathlig@checksuffix@{#1}#2\mathlig@end\fi
}
\def\mathlig@checksuffix@#1#2#3\mathlig@end{%
\expandafter\ifx\csname mathlig@forw@#1#2\endcsname\relax\mathlig@dosuffix{#1}{#2}\fi
\mathlig@checksuffix{#1#2}#3\mathlig@end
}

\def\mathlig@dosuffix#1#2{%
\mathlig@appendcs{mathlig@toks@#1}{#2}%
\mathlig@dolig{#1}{#2}\mathlig@delim
}

\def\mathlig@dolig#1#2\mathlig@delim{%
 \mathlig@defcs{mathlig@look@#1#2}{%
 \mathlig@let@cs\mathlig@next{mathlig@forw@#1#2}\futurelet\mathlig@next@tok\mathlig@next}%
 \mathlig@defcs{mathlig@forw@#1#2}{%
  \mathlig@let@cs\mathlig@next{mathlig@back@#1#2}%
  \mathlig@let@cs\checker{mathlig@chck@#1#2}%
  \mathlig@let@cs\mathligtoks{mathlig@toks@#1#2}%
  \expandafter\ifx\expandafter\mathlig@delim\mathligtoks\mathlig@delim\relax\else
  \expandafter\checker\mathligtoks\mathlig@delim\fi
  \mathlig@next
 }%
 \mathlig@defcs{mathlig@toks@#1#2}{}%
 \mathlig@defcs{mathlig@chck@#1#2}##1##2\mathlig@delim{%
  \ifx\mathlig@next@tok##1%
   \mathlig@let@cs\mathlig@next@cmd{mathlig@look@#1#2##1}\let\mathlig@next\mathlig@gobble
  \fi 
  \ifx\mathlig@delim##2\mathlig@delim\relax\else
   \csname mathlig@chck@#1#2\endcsname##2\mathlig@delim
  \fi
 }%
 \ifx\mathlig@delim#2\mathlig@delim\else
  \mathlig@defcs{mathlig@back@#1#2}{\csname mathlig@back@#1\endcsname #2}%
 \fi
}%

\catcode`@\mathlig@atcode

\usepackage{mathrsfs}
\newcommand{\muspace}{\mspace{1mu}}

\DeclareRobustCommand{\scond}{\mathchoice{\muspace\vert\muspace}{\vert}{\vert}{\vert}}
\mathlig{|}{\scond}
\newcommand{\cond}{\mathchoice{\,\vert\,}{\mspace{2mu}\vert\mspace{2mu}}{\vert}{\vert}}

\DeclareRobustCommand{\discint}{\mathchoice{\mspace{-1.5mu}:\mspace{-1.5mu}}{\mspace{-1.5mu}:\mspace{-1.5mu}}{:}{:}}
\mathlig{::}{\discint}

\def\P{{\mathsf P}}

\newcommand{\Yh}{{\hat{Y}}}

\newcommand{\yh}{{\hat{y}}}

\def\a{\alpha}

\def\eps{\epsilon}

\let\P\relax
\DeclareMathOperator\P{\textsf{P}}

\newcommand{\Bern}{\mathrm{Bern}}

\def\textiid{i.i.d.\@\xspace}
\newcommand\iid{\ifmmode\text{ i.i.d. } \else \textiid \fi}

\def\mathllap{\mathpalette\mathllapinternal}
\def\mathllapinternal#1#2{%
  \llap{$\mathsurround=0pt#1{#2}$}}

\def\clap#1{\hbox to 0pt{\hss#1\hss}}
\def\mathclap{\mathpalette\mathclapinternal}
\def\mathclapinternal#1#2{%
  \clap{$\mathsurround=0pt#1{#2}$}}

\let\oldstackrel\stackrel
\renewcommand{\stackrel}[2]{\oldstackrel{\mathclap{#1}}{#2}}

\renewcommand{\hbar}{h\mathllap{\overline{\vphantom{h}\hphantom{\rule{4.6pt}{0pt}}}\mspace{0.77mu}}}

\catcode`~=11 % make LaTeX treat tilde (~) like a normal character
\newcommand{\urltilde}{\kern -.06em\lower -.06em\hbox{~}\kern .02em}
\catcode`~=13 % revert back to treating tilde (~) as an active character

\begin{document}
\allowdisplaybreaks
\title{Low-Complexity Coding Techniques for Cloud Radio Access Networks}

\author{Nadim Ghaddar, \IEEEmembership{Member, IEEE}, and Lele Wang, \IEEEmembership{Member, IEEE}

\thanks{This work was supported in part by the NSERC Discovery Grant No. RGPIN-2019-05448 and the NSERC CRD Grant CRDPJ 54367619.}

\thanks{Nadim Ghaddar is with the Department of Electrical and Computer Engineering, University of Toronto, Toronto, ON M5S 3G8, Canada (e-mail: nadim.ghaddar@utoronto.ca).
}
\thanks{Lele Wang is with the Department of Electrical and Computer Engineering, University of British Columbia, Vancouver, BC V6T 1Z4, Canada (e-mail: lelewang@ece.ubc.ca).}}

\maketitle

\vspace*{-3em}
\begin{abstract}
The problem of coding for the uplink and downlink of cloud radio access networks (C-RAN's) with $K$ users and $L$ relays is considered. It is shown that low-complexity coding schemes that achieve any point in the rate-fronthaul region of joint coding and compression can be constructed starting from at most $4(K+L)-2$ point-to-point codes designed for symmetric channels. This reduces the seemingly hard task of constructing good codes for C-RAN's to the much better understood task of finding good codes for single-user channels. To show this result, an equivalence between the achievable rate-fronthaul regions of joint coding and successive coding is established. Then, rate-splitting and quantization-splitting techniques are used to show that the task of achieving a rate-fronthaul point in the joint coding region can be simplified to that of achieving a corner point in a higher-dimensional C-RAN problem. As a by-product, some interesting properties of the rate-fronthaul region of joint decoding for uplink C-RAN's are also derived.
\end{abstract}


\section{Introduction} \label{sec:intro}
\begin{figure}[t]
            \vspace{-2em}
    %\begin{multicols}{2}
        \begin{subfigure}[t]{\columnwidth}
            % \centering
            %\hspace*{-8em}
            \footnotesize
            \hspace*{2em}
            \def\svgscale{0.95}
            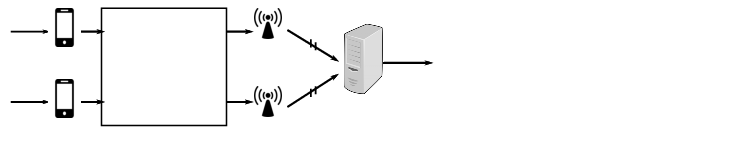
            \caption{}
            \label{fig:uplink_cran_code}
        \end{subfigure}
        \begin{subfigure}[t]{\columnwidth}
            % \centering
            \footnotesize
            \hspace*{1em}
            % \vspace{2em}
            \def\svgscale{0.95}
            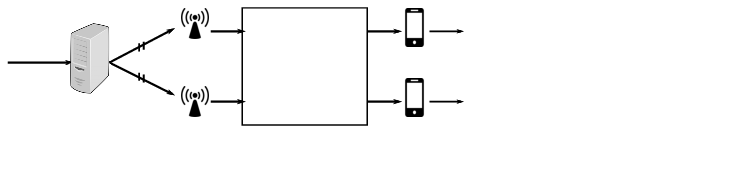
            \caption{}
            \label{fig:downlink_cran_code}
        \end{subfigure}
    %\end{multicols}

    \caption{(a) Uplink and (b) Downlink of a two-user, two-relay C-RAN.}
        \vspace*{-1em}
    \label{fig:codes}
\end{figure}

A cloud radio access network (C-RAN) is an emerging mobile network architecture for next-generation wireless communication systems, promising higher data rates, better coverage and more reliable connectivity for a large number of devices~\cite{Simeone2016}. This architecture has played an important role in the standardization of 5G systems, and is on its way for wide deployment~\cite{Johnson2019,3gpp.38.874}. In a C-RAN architecture, communication between different users is coordinated by a cloud-based central processor, that is connected to the base stations (i.e., relays) of the network via noiseless capacity-limited wired or wireless links. Fig.~\ref{fig:codes} shows schematically the uplink and downlink of a two-user, two-relay cloud radio access network. In the uplink, the users send their encoded messages through the wireless channel, and the relays compress the observations through the fronthaul links to the central processor. In the downlink, the central processor has to compress its transmitted sequences to the relays, which then broadcast the sequences to the users through the channel.

Typically, the users in an uplink scenario encode their messages without utilizing the knowledge of the network structure, and each relay digitizes its received signal individually according to the link capacity constraint. Similarly, in a downlink scenario, the central processor typically performs beamforming assuming that there are no capacity constraints in the fronthaul links, and then the corresponding baseband signals are digitized individually and transmitted to the relays. As an alternative to this greedy approach, this paper investigates coding schemes for the C-RAN architecture by viewing the entire system as a two-hop relay network. 

\vspace{-.5em}
\subsection{Related Work}
In this spirit, coding strategies that utilize the network structure of the C-RAN architecture have been proposed in the literature. Several of these strategies explore the opportunities of either jointly performing the coding and compression functionalities, versus performing the two successively. For the uplink C-RAN model, the compress-forward relaying scheme with joint decoding~\cite{Kramer2005} has been considered in~\cite{Zhou2014, Sanderovich2009, Sanderovich2009_2, Ganguly2021}, where its achievable rate-fronthaul region has been characterized. The authors in~\cite{Aguerri2019} specialized the noisy network coding scheme~\cite{SungHoonLim2011} to uplink C-RAN's and showed that the achievable rate-fronthaul region coincides with the compress-forward relaying scheme with joint decoding. The compress-forward strategy with the more practical successive decoder has been considered in~\cite{Sanderovich2008, Zhou2016}. In particular, in~\cite{Zhou2016}, it is shown that successive decoding can achieve the same rate region as joint decoding under a sum fronthaul capacity constraint. 

For the downlink setting, joint encoding at the relays using linear beamforming has been considered in~\cite{Dai2014, Zakhour2011}, where individual user messages are sent directly through the fronthaul links to the relays. Coding strategies based on Marton's coding scheme for broadcast channels~\cite{Marton1979} have been also employed for downlink C-RAN's~\cite{Yi2015,Wang2018}. In particular,~\cite{Ganguly2021} specializes the distributed decode-forward relaying scheme of a general relay network~\cite{SungHoonLim2017} to the downlink C-RAN problem, where the encoding and compression at the central processor are done jointly. It is shown that the scheme can achieve the capacity region of the Gaussian network to within a constant gap that is logarithmic in the number of users and relays. Alternatively, coding strategies that perform the encoding of user messages and the compression of the analog signals successively have been considered in~\cite{Park2014,Patil2019}, where it is shown that successive encoding can achieve the same rate region as joint encoding when only the sum-capacity of the fronthaul links is constrained.

\vspace*{-1em}
\subsection{Main Contributions}
This paper investigates the relationship between joint decoding (encoding) and successive decoding (encoding) for the uplink (downlink) C-RAN problem. For the uplink case, we show that the compress-forward strategy with successive decoding, where the user messages and quantization codewords can be decoded in an arbitrary order at the central processor, achieves the same rate-fronthaul region as that of joint decoding. Previously, this was shown only for the case of a sum fronthaul capacity constraint~\cite[Theorem 1]{Zhou2016}. Similarly, for the downlink case, we show that a generalized successive encoding strategy, which allows to employ arbitrary encoding orders at the central processor, can achieve the same rate-fronthaul region as the joint encoding strategy (i.e., the distributed decode-forward strategy). Previously, this was shown only for a sum fronthaul constraint~\cite[Theorem 3]{Patil2019}. Motivated by the fact that successive coding is much easier to implement compared to joint coding, these results provide the necessary justification for the implementation of successive coding for the downlink and uplink C-RAN problems.

The successive coding strategies allow to design coding schemes for the C-RAN problem that only use point-to-point codes with single-user encoders and decoders. In our recent work~\cite{Ghaddar2021, Ghaddar2022, Ghaddar2024}, coding schemes for the $K$-user, $L$-relay uplink (downlink) C-RAN problem that achieves a corner point in the joint decoding (encoding) region were constructed starting from $2(K+L)$ point-to-point codes that are designed for symmetric channels. This allows one to leverage the recent advances in point-to-point coding theory over the past few decades (such as list decoding of algebraic codes~\cite{Guruswami1999,Vardy2003}, belief propagation decoding of low-density parity-check codes~\cite{Gallager1962,Mackay1996,Urbanke2013}, and successive cancellation decoding of polar codes~\cite{Arikan2009}) in the construction of low-complexity and rate-optimal coding schemes for the C-RAN problem. Along with the result of this paper highlighted in the previous paragraph, this says that, in fact, one can achieve \emph{any} corner point in the joint decoding (encoding) region of the $K$-user, $L$-relay uplink (downlink) C-RAN problem using $2(K+L)$ point-to-point symmetric channel codes.

This paper also proposes a technique to achieve any point (not just the corner points) in the rate-fronthaul region of joint decoding (encoding) for the uplink (downlink) C-RAN problem without any time sharing. This is particularly useful in the context of the C-RAN problem, where time sharing between corner points with different fronthaul capacity constraints is not possible. This is the case because the fronthaul constraints are not code parameters, but rather are given by the problem statement. Hence, one cannot perform time sharing using distinct codes with rates that satisfy different fronthaul constraints. Even with the restriction to the rate coordinates, the Fenchel--Eggleston--Carath{\'e}odory theorem~\cite[Appendix A]{NIT} asserts that achieving a rate point in the $K$-dimensional space requires time sharing among up to $K+1$ corner points. Hence, this would require up to $2(K+1)(K+L)$ point-to-point codes, potentially with different block lengths. In contrast, the technique that we propose here is based on the notions of rate splitting (introduced by Grant, Rimoldi, Urbanke and Whiting in the context of multiple-access channels~\cite{Grant2001}) and quantization splitting (introduced by Chen and Berger in the context of distributed lossy compression~\cite{Chen2008}), which allow to model the problem of achieving any rate-fronthaul point in the joint coding region as that of coding for a $(2K-1)$-user, $(2L)$-relay C-RAN problem. It follows that $4(K+L)-2$ point-to-point codes are sufficient to construct a coding scheme for the uplink (downlink) C-RAN problem that can achieve any rate-fronthaul point in the joint decoding (encoding) region. 

In summary, this paper makes the following contributions:
\begin{enumerate}
    \item We establish the equivalence between the rate-fronthaul region that can be achieved by joint decoding (encoding) and successive decoding (encoding) for the uplink (downlink) C-RAN problem.
    \item We show that any point on the dominant face of a joint decoding/encoding rate-fronthaul region can be achieved by a splitting scheme with at most $2(K+L)-1$ splits.
    \item We leverage the recently developed Lego-brick approach to show that this task can be accomplished by using at most $4(K+L)-2$ point-to-point codes that are designed for symmetric channels. In other words, the task of constructing good codes for the $K$-user $L$-relay uplink and downlink C-RAN problems can be reduced to that of constructing good codes for at most $4(K+L)-2$ point-to-point symmetric channels.
\end{enumerate}

\vspace*{-1em}
\subsection{Paper Organization and Notation}
The rest of the paper is organized as follows. In Section~\ref{sec:preliminaries}, we define the uplink and downlink C-RAN problems, and show the achievable rate-fronthaul regions of joint and successive coding for both problems. Sections~\ref{sec:equivalence_uplink} and~\ref{sec:p2p_uplink} focus on the uplink C-RAN problem. In Section~\ref{sec:equivalence_uplink}, we show that the rate-fronthaul regions achieved by joint decoding and successive decoding are equivalent, whereas in Section~\ref{sec:p2p_uplink}, we show that any point on the dominant face of the joint decoding rate region can be achieved using point-to-point codes that are designed for symmetric channels. Section~\ref{sec:downlink} briefly discusses analogous results for downlink C-RAN. Section~\ref{sec:conclusion} concludes the paper. The proofs of lemmas are deferred to Appendices.

Notation: We write $x_1^n$ to denote a column vector $(x_1, \ldots, x_n)^T$. 
% Sets are denoted in calligraphic letters, random variables in upper-case letters, and realizations of random variables in lower-case letters, i.e., $\cX$ and $\cY$ are two sets, $X$ and $Y$ are random variables, and $x$ and $y$ are their respective realizations. For a positive integer $n$, we use $[n]$ to denote the set $\{1,\ldots,n\}$. The distribution of a random variable $X$ will be denoted as $p_X = p_X(x)$, $x \in \cX$. The probability of an event $A$ is denoted as $\P\{A\}$. For a finite set $\cS$, we write $\abs{\cS}$ to denote the cardinality of $\cS$. $H(X)$ denotes the entropy of a random variable $X$, and $I(X;Y)$ denotes the mutual information of two random variables $X$ and $Y$. 
Given a vector $R \in \mathbb{R}^n$ and a set $\cS \subseteq [n]$, we write $R_\cS = (R_i: i\in \cS)$ and $R(\cS) = \sum_{i\in \cS}R_i$.

\section{Preliminaries} \label{sec:preliminaries}
\subsection{Uplink C-RAN Problem} \label{sec:preliminaries_uplink}
Consider the uplink of a cloud radio access network with $K$ users and $L$ relays, as shown in Fig.~\ref{fig:uplink_cran_code}. The users wish to communicate with a central processor through the relays that are connected to the central processor through noiseless fronthaul links of finite capacities $C_1, \ldots, C_L$. A memoryless channel $p(y_1^L \cond x_1^K)$ is assumed between the users and the relays, with an input alphabet $\cX_1\times \cdots \times \cX_K$ and output alphabet $\cY_1 \times \cdots \times \cY_L$. An $(R_1,\ldots,R_K,n)$ code for the uplink C-RAN problem consists of
\begin{itemize}
	\item message sets $\cM_1, \ldots, \cM_K$ with $\abs{\cM_k} = 2^{nR_k}, k\in [K]$,
	\item index sets $\cS_1, \ldots, \cS_L$ with $\abs{\cS_\ell} = 2^{nC_\ell}, \ell\in [L]$,
	\item encoders $f_k:\cM_k \to \cX_k^n$ at the $k$-th user that map each message $m_k$ to a codeword $x_k^n$,
	\item encoders $g_\ell: \cY_\ell^n \to \cS_\ell$ at the $\ell$-th relay that map each received sequence $y_\ell^n$ to an index $s_\ell$,
	\item decoder $\psi: \cS_1 \times \cdots \times \cS_M \to \cM_1 \times \cdots \times \cM_K$ at the central processor that assign message estimates $(\hat{m}_1,\ldots, \hat{m}_K)$ to each index tuple $(s_1,\ldots,s_L)$.
\end{itemize}
The average probability of error of the code is defined as $\eps = \P\left\{ \widehat{M}_k \neq M_k \text{ for some } k \in [K] \right\}$. A rate tuple $(R_1,\ldots, R_K)$ is said to be achievable for the uplink C-RAN problem if there exists a sequence of $(R_1,\ldots,R_K,n)$ codes with vanishing error probability asymptotically.

\vspace{0.25em}
\subsubsection{Joint Decoding Rate Region}
The network compress-forward relaying scheme~\cite{Kramer2005} with joint decoding can be specialized to the uplink C-RAN problem~\cite{Zhou2014}, where the user messages and the quantization codewords are decoded jointly at the central processor. In~\cite{Zhou2014}, it is shown that the achievable rate region $\cR_{JD}$ by joint decoding can be expressed as the closure of the convex hull of the set of rate tuples $(R_1,\ldots, R_K,C_1,\ldots,C_L) \in \mathbb{R}_{+}^{K+L}$ satisfying
\begin{equation} \label{eqn:rate_uplink_joint}
	C(\cT) - R(\cS) \geq  I(Y_\cT;\Yh_\cT \cond X_1,\ldots, X_K) - I(X_\cS; \Yh_{\cT^c} \cond X_{\cS^c})
\end{equation}
for all $\cS \subseteq [K]$ and $\cT \subseteq [L]$, for some product distribution $\prod_{k=1}^{K}p(x_k)\prod_{\ell=1}^{L}p(\hat{y}_\ell \cond y_\ell)$.

\vspace{0.25em}
\subsubsection{Successive Decoding Rate Region}
Alternativaly, the compress-forward relaying scheme for the uplink C-RAN problem can be implemented with successive decoding~\cite{Zhou2016}, where the central processor alternates between decoding user messages and quantization codewords according to some prescribed order. More precisely, let $\pi$ be a permutation of $(X_1,\ldots, X_K,\hat{Y}_1,\ldots, \hat{Y}_L)$ corresponding to the decoding order at the central processor. For $k\in [K]$ and $\ell \in [L]$, define
\begin{equation} \label{eqn:index_set}
	\begin{aligned}
		\cI_{X_k} &= \{i \in [K]: X_i \text{ appears before } X_k \text{ under } \pi\},\\
		\cI_{\Yh_\ell} &= \{i \in [K]: X_i \text{ appears before } \Yh_\ell \text{ under } \pi\},\\
		\cJ_{X_k} &= \{j \in [L]: \Yh_j \text{ appears before } X_k \text{ under } \pi\},\\
		\cJ_{\Yh_\ell} &= \{j \in [L]: \Yh_j \text{ appears before } \Yh_\ell \text{ under } \pi\}.
	\end{aligned}
\end{equation}
Rate region $\cR_{SD}\left(\pi\right)$, achieved by the decoding order $\pi$, is the closure of the convex hull of all rate tuples $(R_1,\ldots, R_K,C_1,\ldots,C_L)$ satisfying
\begin{equation} \label{eqn:rate_uplink_successive}
	\begin{cases}
		R_{k} &\hspace*{-0.5em}\leq I(X_{k}; \Yh_{\cJ_{X_k}}, X_{\cI_{X_k}}) \qquad\qquad \quad\;\;\; \forall \, k \in [K],\\
		C_{\ell} &\hspace*{-0.5em}\geq I(Y_{\ell}; \Yh_{\ell}) - I(\Yh_{\ell}; \Yh_{\cJ_{\Yh_\ell}}, X_{\cI_{\Yh_\ell}}) \qquad \forall \, \ell \in [L].
	\end{cases}
\end{equation}
Achievable rate region $\cR_{SD}$ by successive decoding is the closure of the convex hull of the union of rate regions $\cR_{SD}\left(\pi\right)$ over all permutations $\pi$, i.e., $\cR_{SD} = \mathrm{co}\left(\bigcup_{\pi} \cR_{SD}\left(\pi\right)\right)$.
% \[
% \cR_{SD} = \mathrm{co}\left(\bigcup_{\pi} \cR_{SD}\left(\pi\right)\right).
% \]

\vspace{-1em}
\subsection{Downlink C-RAN Problem} \label{sec:preliminaries_downlink}
Consider the downlink of a cloud radio access network with $K$ users and $L$ relays, as shown in Fig.~\ref{fig:downlink_cran_code}. The central processor communicates with the relays through noiseless fronthaul links of finite capacities $C_1, \ldots, C_L$. As in the uplink setting, a memoryless channel $p(y_1^K \cond x_1^L)$ is assumed between the relays and the users, with an input alphabet $\cX_1\times\cdots\times\cX_L$ and output alphabet $\cY_1 \times \cdots \times \cY_K$. An $(R_1,\ldots,R_K,n)$ code for the downlink C-RAN problem consists of
\begin{itemize}
    \item message sets $\cM_1, \ldots, \cM_K$ with $\abs{\cM_k} = 2^{nR_k}, k\in [K]$,
    \item index sets $\cS_1, \ldots, \cS_L$ with $\abs{\cS_\ell} = 2^{nC_\ell}, \ell\in [L]$,
	\item an encoder $f:\cM_1 \times \cdots \times \cM_K \to \cS_1 \times \cdots \times \cS_L$ at the central processor that maps each $(m_1,\ldots,m_K)$ to the indices $(s_1,\ldots,s_L)$ to be sent to the relays,
	\item encoders $g_\ell: \cS_\ell \to \cX_\ell^n$ at the $\ell$-th relay, that map each index $s_\ell$ to a codeword $x_\ell^n$,
	\item decoders $\psi_k: \cY_k^n \to \cM_k$ at the $k$-th user, that assign message estimate $\hat{m}_k$ to each received sequence $y_k^n$.
\end{itemize}
The average probability of error of the code is $\eps = \P\left\{ \widehat{M}_k \neq M_k \text{ for some } k \in [K] \right\}$. A rate tuple $(R_1, \ldots, R_K)$ is said to be achievable for the downlink C-RAN problem if there exists a sequence of $(R_1,R_2,n)$ codes with vanishing error probability asymptotically.

\subsubsection{Joint Encoding Rate Region}
The distributed decode-forward relaying scheme of a general relay network~\cite{SungHoonLim2017}, referred to hereafter as the ``joint encoding'' strategy, has been specialized to the downlink C-RAN problem in~\cite{Ganguly2021}, in which the encoding of user messages and the compression of transmitted codewords are done jointly at the central processor. It is shown that the achievable rate region $\cR_{JE}$ by the joint encoding strategy is the closure of the convex hull of the set of rate tuples $(R_1,\ldots, R_K,C_1,\ldots,C_L) \in \mathbb{R}_{+}^{K+L}$ satisfying
\begin{equation} \label{eqn:rate_downlink_ddf}
	C(\cT) - R(\cS) \geq  I(U_\cS;X_\cT) - \textstyle{\sum_{k \in \cS}} I(U_k;Y_k) + I^{*}(U_\cS)  + I^{*}(X_\cT)
\end{equation}
for all $\cS \subseteq [K]$ and $\cT \subseteq [L]$, and for some distribution $p(u_1,\ldots,u_K,x_1,\ldots,x_L)$, where
\begin{align*}
    I^{*}(U_\cS) &\triangleq \textstyle{\sum_{j \in \cS}}H(U_j) - H(U_\cS),\\
    I^{*}(X_\cT) &\triangleq \textstyle{\sum_{j \in \cT}}H(X_j) - H(X_\cT).
\end{align*}

\subsubsection{Successive Encoding Rate Region}
The successive encoding and compression at the central processor has been considered in~\cite{Patil2019}, where joint encoding of the user messages was followed by joint multivariate compression of the signals intended to relays. We generalize this scheme by allowing the central processor to employ different encoding/compression orders within individual user messages/relay signals. More precisely, let $\pi$ be a permutation of $(U_1,\ldots, U_K,X_1,\ldots, X_L)$ corresponding to the encoding order at the central processor, where $(U_1,\ldots, U_K)$ represent the encoded user messages and $(X_1,\ldots,X_L)$ represent the compressed signals to the relays. For $k\in [K]$ and $\ell \in [L]$, define
\begin{equation} \label{eqn:index_set_downlink}
	\begin{aligned}
		\cI_{U_k} &= \{i \in [K]: U_i \text{ appears before } U_k \text{ under } \pi\},\\
		\cI_{X_\ell} &= \{i \in [K]: U_i \text{ appears before } X_\ell \text{ under } \pi\},\\
		\cJ_{U_k} &= \{j \in [L]: X_j \text{ appears before } U_k \text{ under } \pi\},\\
		\cJ_{X_\ell} &= \{j \in [L]: X_j \text{ appears before } X_\ell \text{ under } \pi\}.
	\end{aligned}
\end{equation}
Rate region $\cR_{SE}\left(\pi\right)$, achieved by the encoding order $\pi$, is the closure of the convex hull of all rate tuples $(R_1,\ldots, R_K,C_1,\ldots,C_L)$ satisfying
\begin{equation} \label{eqn:rate_downlink_gcs}
	\begin{cases}
		R_k \, \leq \, I(U_k;Y_k) - I(U_k;U_{\cI_{U_k}}, X_{\cJ_{U_k}}) & \forall \, k \in [K],\\
		C_\ell \, \geq \, I(X_\ell;U_{\cI_{X_\ell}}, X_{\cJ_{X_\ell}}) & \forall \, \ell \in [L].
	\end{cases}
\end{equation}
Achievable rate region $\cR_{SE}$ by successive encoding is the closure of the convex hull of the union of rate regions $\cR_{SE}\left(\pi\right)$ over all permutations $\pi$, i.e., $\cR_{SE} = \mathrm{co}\left(\bigcup_{\pi} \cR_{SE}\left(\pi\right)\right)$.
% \[
% \cR_{SE} = \mathrm{co}\left(\bigcup_{\pi} \cR_{SE}\left(\pi\right)\right).
% \] 

\section{Equivalence of Joint and Successive Decoding for Uplink C-RAN's} \label{sec:equivalence_uplink}
In this part, we show the equivalence of the rate regions of joint decoding and successive decoding for the uplink C-RAN problem. More specifically, we prove the following theorem.

\begin{theorem} \label{thm:equivalence-uplink}
    The rate region of network compress-forward with joint decoding is equivalent to that of network compress-forward with generalized successive decoding, i.e.,
	\[
	\cR_{JD} = \cR_{SD}.
	\]
\end{theorem}

\noindent Theorem~\ref{thm:equivalence-uplink} asserts that $\cR_{JD}$ has $(K+L)!$ vertices, each of which can be attained by successive decoding of the user messages and quantization codewords according to some prescribed order. To prove Theorem~\ref{thm:equivalence-uplink}, we describe a procedure to find the corner points of the rate region $\cR_{JD}$. This can be done by finding the intersection points of the hyperplanes defined by the set of inequalities in~(\ref{eqn:rate_uplink_joint}). Towards this end, let $S^{K+L}$ be a permutation of $(R_1,\ldots, R_K, C_1,\ldots, C_L)$. We will derive the coordinates of a corner point of $\cR_{JD}$ in the order given by $S^{K+L}$. Define
\begin{equation} \label{eqn:IkJk}
	\begin{aligned}
		\cI_{k} &= \{i \in [K]: S_\ell = R_i \text{ for some } \ell\in [k-1] \},\\
		\cJ_{k} &= \{j \in [L]: S_\ell = C_j \text{ for some } \ell\in [k-1] \}.
	\end{aligned}
\end{equation}

\begin{figure*}
\vspace{-1.5em}
    \noindent \fbox{\parbox{\textwidth}{\textbf{Iterative procedure:} Given a permutation $S^{K+L}$ of $(R_1,\ldots, R_K, C_1,\ldots, C_L)$, for each $k\in [K+L]$, set
    \begin{equation} \label{eqn:procedure_iterative}
    	S_k = \begin{cases}
    		C(\cJ_k) - R(\cI_k) - I(Y_{\cJ_k};\hat{Y}_{\cJ_k} \cond X_1,\ldots, X_K)  + I(X_{b_k}, X_{\cI_k}; \hat{Y}_{\cJ_k^c} \cond X_{\cI_k^c\setminus \{b_k\}}) &\hspace*{-0.35em}\text{ if } a_k = 1,\\
    		R(\cI_k) - C(\cJ_k) +  I(Y_{b_k}, Y_{\cJ_k};\hat{Y}_{b_k}, \hat{Y}_{\cJ_k} \cond X_1,\ldots, X_K) - I(X_{\cI_k};\hat{Y}_{\cJ_k^c\setminus \{b_k\}} \cond X_{\cI_k^c}) &\hspace*{-0.35em}\text{ if } a_k = 0.
    	\end{cases}
    \end{equation}
    }}
    \vspace{-1.5em}
\end{figure*}

\noindent As an example, if $K=3$, $L=2$, and $S^{K+L} = (R_3, R_1, C_2, R_2, C_1)$, then we have $\cI_1 = \emptyset, \cI_2 = \{3\}, \cI_3 = \cI_4 = \{1,3\},  \cI_5 = \{1,2,3\}, \cJ_1 = \cJ_2 =  \cJ_3 = \emptyset, \cJ_4 = \cJ_5 = \{2\}.$
% \begin{align*}
% 	\cI_1 = \emptyset \quad &\text{ and } \quad \cJ_1 = \emptyset,\\
% 	\cI_2 = \{3\} \quad &\text{ and } \quad \cJ_2 = \emptyset,\\
% 	\cI_3 = \{1,3\} \quad &\text{ and } \quad \cJ_3 = \emptyset,\\
% 	\cI_4 = \{1,3\} \quad &\text{ and } \quad \cJ_4 = \{2\},\\
% 	\cI_5 = \{1,2,3\} \quad &\text{ and } \quad \cJ_5 = \{2\}.
% \end{align*}
Moreover, let $a^{K+L} = (a_1,\ldots,a_{K+L})$ and $b^{K+L} = (b_1,\ldots,b_{K+L})$ be defined such that
\begin{equation} \label{eqn:akbk}
	\begin{aligned}
		a_k &= \begin{cases}
			1 &\qquad \text{ if } S_k = R_i \text{ for some } i \in [K],\\
			0 &\qquad \text{ if } S_k = C_j \text{ for some } j \in [L],
		\end{cases}\\
		b_k &= \begin{cases}
			i &\qquad \text{ if } a_k = 1 \text{ and } S_k = R_i,\\
			j &\qquad \text{ if } a_k = 0 \text{ and } S_k = C_j.
		\end{cases}
	\end{aligned}
\end{equation}
for each $k \in [K+L]$. Therefore, if $a_k=1$, we have $S_k = R_{b_k}$, and if $a_k=0$, we have $S_k = C_{b_k}$. For the example considered previously, $a^{K+L} = (1,1,0,1,0)$ and $b^{K+L} = (3,1,2,2,1)$. 

Now, consider the iterative procedure for setting $S^{K+L}$ described by equation~(\ref{eqn:procedure_iterative}) at the top of the next page. Note that this procedure corresponds to setting the inequality in~(\ref{eqn:rate_uplink_joint}) to equality, with
\begin{equation} \label{eqn:set_choice}
	\begin{cases}
		\cS = \cI_k \cup \{b_k\} \text{ and } \cT = \cJ_k &\text{ if } a_k=1,\\
		\cS = \cI_k \text{ and } \cT = \cJ_k \cup \{b_k\} &\text{ if } a_k=0.
	\end{cases}
\end{equation}

\begin{lemma} \label{lemma:procedure_iterative_modified}
	The procedure described by equations~(\ref{eqn:procedure_iterative}) is equivalent to setting, 	for each $k \in [K+L]$,
	\begin{equation} \label{eqn:procedure_iterative_modified}
\hspace{-.5em}		S_k = \begin{cases}
			I(X_{b_k}; \hat{Y}_{\cJ_k^c} \cond X_{\cI_k^c \setminus \{b_k\}})& \!\!\!\!\text{if } a_k = 1,\\
			I(Y_{b_k}; \hat{Y}_{b_k}) - I(\hat{Y}_{b_k}; X_{\cI_k^c}, \hat{Y}_{\cJ_k^c \setminus \{b_k\}})& \!\!\!\!\text{if } a_k = 0.
		\end{cases}
	\end{equation}

\end{lemma}

% \begin{proof}
%     See Appendix~\ref{appx:procedure_iterative_modified}.
% \end{proof}

\begin{lemma} \label{lemma:corner_point}
    The procedure described by~(\ref{eqn:procedure_iterative}) gives a corner point of the rate region $\cR_{JD}$.
\end{lemma}

% \begin{proof}
%     See Appendix~\ref{appx:corner_point}.
% \end{proof}

Lemmas~\ref{lemma:procedure_iterative_modified} and~\ref{lemma:corner_point} given an explicit way of finding any corner point of the joint decoding rate region $\cR_{JD}$ through the procedure described in~(\ref{eqn:procedure_iterative}). With these lemmas, we are ready to prove Theorem~\ref{thm:equivalence-uplink}. Clearly, we have that $\cR_{SD} \subseteq \cR_{JD}$. To show the other direction, it suffices to show that every corner point of $\cR_{JD}$ belongs to $\cR_{SD}$. From Lemmas~\ref{lemma:procedure_iterative_modified} and~\ref{lemma:corner_point}, we know that any corner point of $\cR_{JD}$ can be expressed iteratively as follows:
\begin{equation} \label{eqn:procedure_modified_2}
 \hspace{-.5em}   S_k = \begin{cases}
        I(X_{b_k}; \hat{Y}_{\cJ_k^c} \cond X_{\cI_k^c \setminus \{b_k\}})&\!\!\!\!\text{if } a_k = 1,\\
        I(Y_{b_k}; \hat{Y}_{b_k}) - I(\hat{Y}_{b_k}; X_{\cI_k^c}, \hat{Y}_{\cJ_k^c \setminus \{b_k\}})&\!\!\!\! \text{if } a_k = 0.
    \end{cases}
\end{equation}
for each $k \in [K+L]$, where $S^{K+L}$ is some permutation of $(R_1,\ldots, R_K, C_1,\ldots, C_L)$, and $(\cI_k,\cJ_k, a_k,b_k)$ are defined as in~(\ref{eqn:IkJk}) and~(\ref{eqn:akbk}). Let $\pi = (\pi_1,\ldots,\pi_{K+L})$ be a permutation of $(X_1,\ldots,X_K, \hat{Y}_1,\ldots, \hat{Y}_L)$ such that for each $k \in [K+L]$,
\begin{equation}
    \pi_k = \begin{cases}
        X_{b_{K+L-k+1}} &\quad \text{ if } a_k = 1,\\
        \hat{Y}_{b_{K+L-k+1}} &\quad \text{ if } a_k = 0.
    \end{cases}
\end{equation}
Comparing expression~(\ref{eqn:procedure_modified_2}) with rate region~(\ref{eqn:rate_uplink_successive}) of successive decoding, it can be seen that the corner point $S^{K+L}$ of $\cR_{JD}$ is also a corner point of $\cR_{SD}$ when decoding order $\pi$ is used. This completes the proof of~Theorem~\ref{thm:equivalence-uplink}.

\section{Achieving the Joint Decoding Rate Region of Uplink C-RAN's Using Point-to-Point Codes} \label{sec:p2p_uplink}
In this section, we show that any point in the rate region of joint decoding for the uplink C-RAN problem can be achieved using point-to-point codes that are designed for symmetric channels\footnote{Recall the definition of a symmetric channel~\cite[Section 4.5]{Gallager1968}. A discrete memoryless channel $p(y \cond x)$ is symmetric if the set of outputs can be partitioned into subsets such that for each subset, the matrix of transition probabilities has the property that the rows are permutations of each other, and the columns are permutations of each other. Note that linear codes can only achieve the capacity of symmetric channels.}. This allows one to leverage commercial off-the-shelf point-to-point codes in the implementation of coding schemes for the uplink C-RAN problem. More precisely, we prove the following theorem.

\begin{theorem} \label{thm:uplink}
    Every rate-fronthaul tuple $(R_1,\ldots,R_K,$ $C_1,\ldots,C_L) \in \cR_{JD}$ can be achieved via a coding scheme that is constructed starting from at most $4(K+L)-2$ point-to-point symmetric channel codes.
\end{theorem}

The proof of Theorem~\ref{thm:uplink} consists of two main parts: (1) showing that every point in the rate-fronthaul region of joint decoding can be achieved via successive decoding over a higher-dimensional C-RAN problem, and (2) leveraging point-to-point channel codes to code for the higher-dimensional problem. To show (1), we first derive some interesting properties of the joint decoding rate region in Section~\ref{sec:dominant_uplink}, which are analogous to the structure of the capacity region of multiple-access channels~\cite{Grant2001}. Then, in Section~\ref{sec:splitting_existing}, we review existing rate-splitting and quantization-splitting techniques to code for the multiple-access channel~\cite{Grant2001} and the distributed lossy compression problem~\cite{Chen2008}. These techniques will be the stepping stones towards showing (1) for the uplink C-RAN problem in Section~\ref{sec:splitting_uplink}. Finally, to show (2), we refer to the Lego-brick framework of~\cite{Ghaddar2024} and describe a coding scheme for the higher-dimensional uplink C-RAN problem using point-to-point channel codes in Section~\ref{sec:p2p_uplink_sub}.

 \vspace{-1em}
\subsection{Dominant Face of Joint Decoding Rate Region} \label{sec:dominant_uplink}
Towards proving Theorem~\ref{thm:uplink}, we first study the structure of the dominant face of the joint decoding rate region.

\begin{definition}
    We say that a rate-fronthaul tuple $(R_1, \ldots, R_K, C_1, \ldots, C_L)$ \emph{dominates} another tuple $(R_1', \ldots, R_K', C_1', \ldots, C_L')$ if and only if $R_i \geq R_i'$ for each $i \in [K]$, and $C_i \leq C_i'$ for each $i \in [L]$.
\end{definition}

\begin{definition}[Dominant Face]
    The \emph{dominant face} $\cD$ of $\cR_{JD}$ is the convex polytope consisting of all rate-fronthaul tuples $(R_1, \ldots, R_K, C_1, \ldots, C_L) \in \cR_{JD}$ such that
    \[
    C([L]) - R([K]) = I(Y_1,\ldots,Y_L; \Yh_1,\ldots, \Yh_L \cond X_1,\ldots, X_K).
    \]
    This corresponds to setting the inequality in~(\ref{eqn:rate_uplink_joint}) to equality when $\cS = [K]$ and $\cT = [L]$. 
\end{definition}

Note that every point in $\cR_{JD}$ is dominated by a point on the dominant face $\cD$. Hence, in what follows, we restrict our attention to rate-fronhaul tuples that belong to $\cD$. The following lemma gives an alternative description of the dominant face $\cD$ which will be useful in the subsequent analysis.

\begin{lemma} \label{lemma:dominant}
    The dominant face $\cD$ can be expressed as     
    $
        \cD = \big\{(R_1, \ldots, R_K, C_1, \ldots, C_L): \forall \, \cS \subseteq [K], \cT \subseteq [L],$ $ 
         I(Y_\cT;\Yh_\cT \cond X_1^K) - I(X_\cS;\Yh_{\cT^c} \cond X_{\cS^c}) \leq C(\cT)-R(\cS) 
        \leq I(Y_\cT;\Yh_\cT \cond X_\cS)\big\}.$
    % \begin{align*}
    %     \cD &= \Big\{(R_1, \ldots, R_K, C_1, \ldots, C_L): \forall \, \cS \subseteq [K], \cT \subseteq [L], \\
    %     &\qquad I(Y_\cT;\Yh_\cT \cond X_1^K) - I(X_\cS;\Yh_{\cT^c} \cond X_{\cS^c}) \leq C(\cT)-R(\cS) \\
    %     &\qquad \leq I(Y_\cT;\Yh_\cT \cond X_\cS)\Big\}.
    % \end{align*}
\end{lemma}
% \begin{proof}
%     See Appendix~\ref{appx:dominant}.
% \end{proof}

Now, we focus on the boundary of $\cD$. For $\cS \subseteq [K]$ and $\cT \subseteq [L]$ such that $\cS^c \cup \cT^c \neq \emptyset$ and $\cS \cup \cT \neq \emptyset$, define %the hyperplane
\begin{align*}
    \cH_{\cS,\cT} = \{&(R_1,\ldots,R_K,C_1,\ldots, C_L): \\
    &C(\cT) - R(\cS) = I(Y_\cT;\Yh_\cT \cond X_\cS)\},
\end{align*}
and let $\cF_{\cS,\cT} = \cD \cap \cH_{\cS,\cT}$. By Lemma~\ref{lemma:dominant}, we know that $\cH_{\cS,\cT}$ lives entirely on one side of $\cD$, and hence, $\cF_{\cS,\cT}$ defines a face of $\cD$. Analogous to the case of multiple-access channels~\cite{Grant2001}, the set $\cF_{\cS,\cT}$ has an interesting operational meaning pertaining to the decoding order at the central processor. Towards the goal of characterizing this meaning, consider the uplink C-RAN problem defined by the channel $W_{\cS,\cT}: \bigotimes_{i\in \cS} \cX_i \to \bigotimes_{i\in \cT} \cY_i$ which can be described by stochastic matrix $W_{\cS,\cT}$ with entries
$
W_{\cS,\cT}(y_{\cT} \cond x_\cS) = \textstyle{\sum_{x_{\cS^c}}\sum_{y_{\cT^c}}} p(x_{\cS^c})p(y_1,\ldots,y_L \cond x_1,\ldots, x_K).
$
This is the channel with inputs indexed by $\cS$ and outputs indexed by $\cT$ when the remaining inputs are treated as ``noise''. Similarly, consider the uplink C-RAN problem defined by the channel $W_{\cS^c,\cT^c \cond \cS,\cT}: \bigotimes_{i\in \cS^c} \cX_i \to \bigotimes_{i\in \cT^c} \cY_i$ which can be described by the stochastic matrix $W_{\cS^c,\cT^c \cond \cS,\cT}$ with entries
$
    W_{\cS^c,\cT^c \cond \cS,\cT}(y_{\cT^c}, x_\cS, \hat{y}_\cT \cond x_{\cS^c})= p(x_{\cS}) p(\hat{y}_\cT \cond x_1^K)p(y_{\cT^c} \cond x_1^K).
$
This is the channel with inputs indexed by $\cS^c$ and outputs indexed by $\cT^c$, when $x_\cS$ and $\hat{y}_\cT$ are available to the central processor as side information. Let $\cD_{\cS,\cT}$ and $\cD_{\cS^c,\cT^c \cond \cS,\cT}$ be the dominant faces associated to joint decoding for the uplink C-RAN problems defined by $W_{\cS,\cT}$ and $W_{\cS^c,\cT^c \cond \cS,\cT}$, respectively. Explicitly, one can show that % \nadim{[Show why that is the case]}
\begin{align*}
    &\cD_{\cS,\cT} = \Big\{(R_\cS,C_\cT) \in \mathbb{R}^{\abs{\cS} + \abs{\cT}}: \forall \, \cA \subseteq \cS, \cB \subseteq \cT, \\
    &\qquad\qquad I(Y_\cB;\Yh_\cB \cond X_\cS, \Yh_{\cT \setminus \cB}) - I(X_\cA;\Yh_{\cT\setminus \cB}\cond X_{\cS\setminus \cA}) \\
    &\qquad\qquad \leq C(\cB) - R(\cA) \leq I(Y_{\cB}; \Yh_{\cB} \cond X_{\cA})\Big\},\\
    &\cD_{\cS^c,\cT^c \cond \cS,\cT} = \Big\{(R_{\cS^c},C_{\cT^c}) \in \mathbb{R}^{\abs{\cS^c} + \abs{\cT^c}}: \forall \, \cA \subseteq \cS^c, \cB \subseteq \cT^c, \\
    &\quad  f_{\cS^c,\cT^c \cond \cS,\cT}^{\mathrm{lb}}(\cA,\cB) \leq C(\cB) - R(\cA) \leq f_{\cS^c,\cT^c \cond \cS,\cT}^{\mathrm{ub}}(\cA,\cB)\Big\},
\end{align*}
where
\begin{align*}
    f_{\cS^c,\cT^c \cond \cS,\cT}^{\mathrm{lb}}(\cA,\cB) = &I(Y_\cB; \Yh_\cB \cond X_{\cS^c}, \Yh_{\cT^c \setminus \cB}, X_\cS, \Yh_\cT) \\
    &- I(X_\cA; \Yh_{\cT^c \setminus \cB}, \Yh_\cT \cond X_{\cS^c \setminus \cA}, X_\cS),\\
    f_{\cS^c,\cT^c \cond \cS,\cT}^{\mathrm{ub}}(\cA,\cB) = &I(Y_{\cB}; \Yh_{\cB} \cond X_\cA, X_{\cS},\Yh_{\cT}) - I(X_\cA; \Yh_\cT \cond  X_\cS).
\end{align*}

\begin{lemma} \label{lemma:boundary}
    For any $\cS \subseteq [K]$ and $\cT \subseteq [L]$ such that $\cS^c \cup \cT^c \neq \emptyset$ and $\cS \cup \cT \neq \emptyset$, we have
    $
    \cF_{\cS,\cT} = \cD_{\cS,\cT} \times \cD_{\cS^c,\cT^c \cond \cS,\cT}.
    $
\end{lemma}

% \begin{proof}
%     See Appendix~\ref{appx:boundary}.
% \end{proof}

Lemma~\ref{lemma:boundary} says that when $(R_1,\ldots,R_K,C_1,\ldots,C_L) \in \cF_{\cS,\cT}$, the subset of user messages in $\cS$ and quantization codewords in $\cT$ can be decoded first (e.g., via a joint decoder that considers inputs from users in $\cS^c$ as noise), and then the user messages in $\cS^c$ and quantization codewords in $\cT^c$ can be decoded by taking the decoded $(x_\cS,\hat{y}_\cT)$ as side information. This insight says that one can employ two joint decoders \emph{successively} to achieve any rate-fronthaul tuple on the boundary of $\cD$. Nonetheless, Theorem~\ref{thm:uplink} says more than this; namely, that one can achieve any rate-fronthaul tuple on the dominant face $\cD$ by successively decoding, one by one, the user messages and quantization codewords. 

\begin{lemma} \label{lemma:degenerate-1}
    For any $\cS \subseteq [K]$ and $\cT \subseteq [L]$ such that $\cS \cup \cT \neq \emptyset$ and $\cS^c \cup \cT^c \neq \emptyset$, the following are equivalent:
    \begin{enumerate}[label={\alph*)}]
        \item $I(X_\cS ; \Yh_{\cT^c} \cond X_{\cS^c}) = 0$ and $I(X_{\cS^c} ; \Yh_{\cT} \cond X_\cS) = 0$;
        \item $\cD = \cD_{\cS,\cT} \times \cD_{\cS^c,\cT^c}$.
    \end{enumerate}
\end{lemma}

% \begin{proof}
%     See Appendix~\ref{appx:degenerate-1}.
% \end{proof}

\begin{lemma} \label{lemma:degenerate-2}
    The following statements are equivalent:
    \begin{enumerate}[label={\alph*)}]
        \item $\cD = \cD_{\cS,\cT}\times \cD_{\cS^c,\cT^c}$ for some $\cS \subseteq [K]$ and $\cT \subseteq [L]$ such that $\cS \cup \cT \neq \emptyset$ and $\cS^c \cup \cT^c \neq \emptyset$;
        \item $\dim(\cD)< K+L-1$.
    \end{enumerate}
\end{lemma}

% \begin{proof}
%     See Appendix~\ref{appx:degenerate-2}.
% \end{proof}

Lemmas~\ref{lemma:degenerate-1} and~\ref{lemma:degenerate-2} give sufficient and necessary conditions for the case when the dimension of the dominant face $\cD$ of $\cR_{JD}$ is strictly less than $K+L-1$. In words, the lemmas say that when $\dim(\cD) < K+L-1$, the uplink C-RAN problem can be decomposed into two separate uplink C-RAN problems with independent underlying channels.

\vspace{-1em}
\subsection{Rate Splitting and Quantization Splitting} \label{sec:splitting_existing}

\begin{figure}[t] {
\vspace{-2em}
    \centering
    \hspace*{-1em}
    \footnotesize
    \def\svgscale{0.95}
    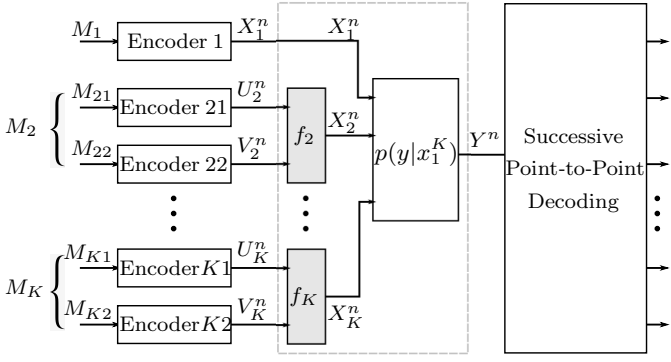
    \caption{Rate splitting for the $K$-user multiple access channel.}
    \label{fig:splitting_mac}
    \vspace{-1em}
}
\end{figure}

Rate splitting was introduced in~\cite{Grant2001} as a technique to achieve any point in the capacity region of the multiple-access channel via single-user decoding methods. The idea is to split each input $X_i$ of the multiple-access channel to two independent ``virtual'' inputs $U_i$ and $V_i$ with some probability mass functions (pmf's) $p_U$ and $p_V$. More precisely, the following definition of a rate split can be seen as a recast of~\cite[Definition 2]{Grant2001}.

\begin{definition}[Rate Split]
    A rate split of a channel input with alphabet $\cX$ and pmf $p_X$ is the set of triples $(f,p_U,p_V)$ parameterized by a real number $\epsilon \in [0,1]$, where $f:\cX \times \cX \to \cX$ is the splitting function, and $p_U$ and $p_V$ are pmf's that depend on $\epsilon$. When $U$ and $V$ are independent random variables with pmf's $p_U$ and $p_V$, the triple $(f,p_U,p_V)$ is such that
    \begin{enumerate}[label={\roman*)}]
        \item $f(U,V)\sim p_X$, regardless of the parameter $\eps$,
        \item $p_{f(U,V) \cond U}(x\cond u)$ is a continuous function of $\epsilon, \forall (u,x)$,
        \item $p_{f(U,V) \cond U} = p_X$ when $\epsilon=0$ (i.e., $f(U,V)$ is independent of $U$), and $p_{f(U,V) \cond U}$ is deterministic when $\epsilon = 1$ (i.e., $f(U,V)$ is completely determined by $U$).
    \end{enumerate}
\end{definition}

\noindent A rate split with the desired properties is shown to exist for finite-input alphabets in~\cite{Grant2001} and for Gaussian channels in~\cite{Rimoldi1996}. Using such a splitting technique on the channel inputs, it is shown in~\cite{Grant2001} that any rate tuple in the capacity region of the $K$-user multiple-access channel can be achieved using at most $2K-1$ rate splits, as shown in Fig.~\ref{fig:splitting_mac}.

\begin{example}
    Consider $\cX = \{0,1\}$, and let $p_X$ be the $\Bern(\alpha)$ distribution. For any parameter $\eps \in [0,1]$, let $p_U$ and $p_V$ be the independent $\Bern(\alpha\eps)$ and $\Bern\left(\frac{\alpha(1-\eps)}{1-\alpha\eps}\right)$ distributions. Let $f(u,v) = \max\{u,v\}$ be the splitting function. When $(U,V) \sim p_Up_V$, it is straightforward to check that $\P(f(U,V) = 1) = \alpha = p_X(1)$. Moreover,
    $
    \P(f(U,V) = 1 \cond U = 0) = \tfrac{\alpha(1-\eps)}{1-\alpha\eps},
    $
    and $\P(f(U,V) = 1 \cond U = 1) = 1$, which are both continuous functions of $\eps$. Finally, $\eps = 0$ implies that $U=0$, and thus, $f(U,V) = V$, which is independent of $U$, whereas $\eps=1$ implies that $V=0$, and thus, $f(U,V)=U$, i.e., $f(U,V)$ is completely determined by $U$.
\end{example}

Quantization splitting is an analogous technique introduced in~\cite{Chen2008} to achieve any tuple in the Berger--Tung rate region of the distributed lossy compression problem via successive Wyner--Ziv coding. The idea is to generate for each source $Y$ two representations $U$ and $V$ which, when combined together, give the same representation as when no splitting is performed. One can view $U$ as a coarse description of the source, which gets finer after observing the other description $V$. Specifically, the following definition of a quantization split can be seen as a recast of the conditions in~\cite[Theorem 2.1]{Chen2008}.

\begin{definition}[Quantization Split]
    A quantization split of a source $Y$ with alphabet $\cY$, source pmf $p_Y$ and desired source-reconstruction pmf $p_{Y,\Yh}$ is a family of pairs $(g,p_{U,V \cond Y})$ parameterized by a real number $\epsilon \in [0,1]$, where $g:\hat{\cY} \times \hat{\cY} \to \hat{\cY}$ is the splitting function, and $p_{U,V \cond Y}$ can depend on $\eps$. When $(U,V)$ are generated conditionally given $Y$ according to $p_{U,V \cond Y}$, the pair $(g,p_{U,V \cond Y})$ satisfies 
    \begin{enumerate}[label={\roman*)}]
        \item $Y - \Yh - (U,V)$ forms a Markov chain for each $\eps$,
        \item $(Y, g(U,V) ) \sim p_{Y,\Yh}$, regardless of the parameter $\eps$,
        \item $p_{g(U,V) \cond U}(\hat{y}\cond u)$ is a continuous function of $\epsilon$ for all $(u,\hat{y})$,
        \item $p_{g(U,V) \cond U} = p_{\Yh}$ when $\epsilon=0$ (i.e., $g(U,V)$ is independent of $U$), and $p_{g(U,V) \cond U}$ is deterministic when $\epsilon = 1$ (i.e., $g(U,V)$ is completely determined by $U$).
    \end{enumerate}
\end{definition}

\begin{figure}[t] {
\vspace{-2em}
    \centering
    \hspace*{-0.5em}
    \footnotesize
    \def\svgscale{1.05}
    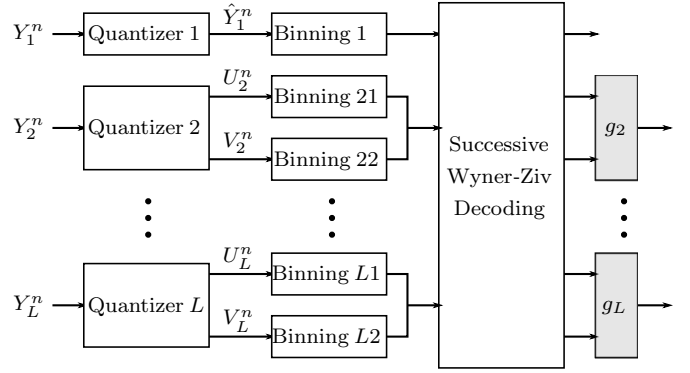
    \caption{Quantization splitting for $L$-source distributed lossy compression.}
    \label{fig:splitting_berger}
    \vspace{-1em}
}
\end{figure}

\noindent A quantization split with the desired properties is shown to exist in~\cite{Chen2008}. Using such a splitting technique along with successive Wyner--Ziv decoding, it is shown that any rate tuple in the Berger--Tung rate region can be achieved (see Fig.~\ref{fig:splitting_berger}).

\begin{example}
    Consider $\cY= \hat{\cY} = \{0,1\}$, and let $p_Y$ be the $\Bern(\alpha)$ distribution, and let $p_{\Yh\cond Y}$ be a $\BSC(\beta)$, the binary symmetric channel with crossover probability $\beta$. For any parameter $\eps \in [0,1]$, we will construct $p_{U,V \cond Y}$ such that property i) holds. More specifically, when $(Y,\Yh) \sim p_{Y,\Yh}$, let $T\sim\Bern(\eps)$ be independent of $(Y,\Yh)$, and then set $(U,V) = (0,\Yh)$ if $T=0$ and $(U,V) = (\Yh,0)$ if $T=1$.
    % \[
    % (U,V) = \begin{cases}
    %     (0,\Yh) &\text{if } T = 0,\\
    %     (\Yh,0) &\text{if } T = 1.
    % \end{cases}
    % \]
    Given $\Yh$, $(U,V)$ is independent of $Y$. Hence, we define
$
    p_{U,V\cond Y}(u,v \cond y) = \textstyle{\sum_{\hat{y}}} p_{\Yh \cond Y} (\hat{y} \cond y)\P(U=u,V=v \cond \Yh=\hat{y}),
$
    for each $u,v,y \in \{0,1\}$, and let $g(u,v) = \max\{u,v\}$ be the splitting function. One can check that properties ii)--iv) hold.
\end{example}

\vspace{-1em}
\subsection{Splitting for the Uplink C-RAN problem} \label{sec:splitting_uplink}
To prove Theorem~\ref{thm:uplink}, we apply a set of quantization splits on the quantization codewords in addition to the set of rate splits on the channel inputs. The proof technique is inspired by the proof of~\cite[Theorem 10]{Grant2001}. Here, to avoid the repetition, we set up the necessary notation for the main ingredient of the proof, and then refer to~\cite{Grant2001} for the details.

\begin{figure*}[t] {
\vspace{-2em}
    \centering
    \hspace*{5.5em}
    \small
    \def\svgscale{0.85}
    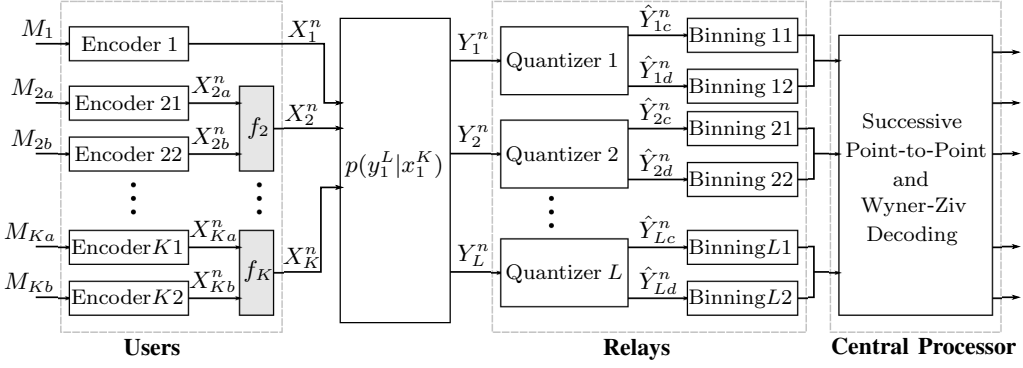
    % \vspace{-.5em}
    \caption{Splitting for the $K$-user $L$-relay uplink C-RAN problem.}
    \label{fig:splitting_uplink_cran}
    \vspace{-1em}
}
\end{figure*}

\begin{figure}[t] {
    \centering
    \hspace*{1.5em}
    \small
    \def\svgscale{0.8}
    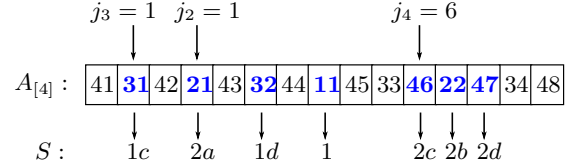
    \caption{Generalized order for $K=2$ and $L=2$. Bold entries correspond to the active elements when $j_2=1$, $j_3 =1$ and $j_4 = 6$.}
    \label{fig:decoding_order}
    \vspace{-1.5em}
}
\end{figure}

Towards this end, consider a product distribution
$
\textstyle{\prod_{i=1}^{K}p(x_i)\prod_{i=1}^{L}p(\hat{y}_i \cond y_i)}
$
for the uplink C-RAN problem, and let $(R_1,\ldots, R_K, C_1 , \ldots, C_L)$ be an arbitrary rate-fronthaul tuple on the dominant face $\cD$ of $\cR_{JD}$ defined by this product distribution. We will show that any such rate-fronthaul tuple can be achieved as shown in Fig.~\ref{fig:splitting_uplink_cran}. Specifically, for $i = 2,\ldots, K$, we split the channel input $X_i$ using a rate split to two virtual inputs $X_{ia}$ and $X_{ib}$, where the channel input $X_1$ remains un-split. Moreover, for $i=1,\ldots, L$, we split the quantization codeword $\Yh_i$ using a quantization split to two virtual representations $\Yh_{ic}$ and $\Yh_{id}$. In this process, we will use a vector of splitting parameters $\eps = (\eps_2,\ldots, \eps_K,\eps_{K+1}, \ldots, \eps_{K+L})$, where $\eps_i$ corresponds to a rate split when $i\in \{2,\ldots, K\}$ and to a quantization split when $i\in \{K+1,\ldots, K+L\}$. To further set up the notation, let
$
\cP = \cP_1 \cup \cP_2,
$
where $\cP_1 = \{1,2a,2b,\ldots, Ka, Kb\}$ and $\cP_2 = \{1c, 1d, \ldots, Lc, Ld\}$ are the sets of indices of virtual inputs and quantization codewords, respectively. Let $S = (S_1,\ldots, S_{2(K+L)-1})$ be a permutation on $\cP$ defining the decoding order of the virtual inputs and quantization codewords at the central processor\footnote{We assume that permutation $S$ is well-ordered in the sense that $ia$ appears before $ib$ for each $i\in [K]\setminus \{1\}$, and $ic$ appears before $id$ for each $i \in [L]$.}. For example, if $K=2$, $L=2$, and $S = (1c,2a,1d,1,2c,2b,2d)$, then the central processor decodes in the order $(\Yh_{1c}, X_{2a}, \Yh_{1d}, X_{1}, \Yh_{2c}, X_{2b}, \Yh_{2d})$. The splitting parameter vector $\eps$ and the decoding order $S$ are chosen as a function of a parameter vector $\alpha = (\alpha_2, \ldots,\alpha_K,\alpha_{K+1}, \ldots, \alpha_{K+L}) \in [0,1]^{K+L-1}$. This is done in the exact same way as in~\cite{Grant2001}. For completion, we provide the details here. For each $i\in [K+L]$, let $m_i = 2^{i-1}-1$. The splitting parameter vector is chosen based on $\alpha$ as follows. For $i\in[K+L]\setminus \{1\}$, let
\begin{align*}
    j_i &= j_i(\alpha_i) = \begin{cases}
        \lfloor \alpha_i m_i \rfloor +1, &\;\text{if } \alpha_i \in [0,1),\\
        m_i, &\;\text{if } \alpha_i =1,
    \end{cases}\\
    \eps_i &= \eps_i(\alpha_i) = \alpha_i m_i - j_i(\alpha_i) + 1.
\end{align*}
This corresponds to dividing the interval $[0,m_i]$ into $m_i$ equal subintervals and choosing $\eps_i$ as the normalized position of $\alpha_i m_i$ within the subinterval containing it. Thus, this defines, for every parameter vector $\alpha$, a product distribution
$
p_{X_1}p_{X_{2a}}p_{X_{2b}}\cdots p_{X_{Ka}}p_{X_{Kb}}p_{\Yh_{1c}\Yh_{1d} \cond Y_1}\cdots p_{\Yh_{Lc}\Yh_{Ld} \cond Y_L}.
$
The decoding order $S$ is then assigned as follows. Let $A_{[K+L]}$ be a generalized order defined recursively. Starting from $A_{[1]} = (11)$, $A_{[i]}$ is obtained by interleaving $A_{[i-1]}$ with $(i1,i2,\ldots,i(2^{i-1}))$. For example, $A_{[2]} = (21,11,22)$ and $A_{[3]} = (31,21,32,11,33,22,34)$. Given the generalized order $A_{[K+L]}$, the elements $ij_i$ and $i(j_i+1)$ are labeled as \emph{active} for each $i \in [K+L]\setminus \{1\}$. The element $11$ is always labeled as active. Then, the decoding order $S$ is assigned by reading the active elements in their order of appearance in $A_{[K+L]}$, while relabeling the elements to their corresponding entry in $\cP$. As an example, Fig.~\ref{fig:decoding_order} shows the assigned decoding order $S$ when $K=2$, $L=2$, and the active indices are $j_2=1$, $j_3 =1$ and $j_4 = 6$. This would correspond to a decoding order as in the example given previously. 

Now, let $a = (a_1,\ldots, a_{2(K+L)-1})$ be such that $a_k = 1$ if $S_k \in \cP_1$ and $a_k = 0$ if $S_k \in \cP_2$,
% \[
% a_k = \begin{cases}
%     1 &\quad \text{if } S_k \in \cP_1,\\
%     0 &\quad \text{if } S_k \in \cP_2,
% \end{cases}
% \]
and let $b = (b_1,\ldots, b_{2(K+L)-1})$ be such that
\[
b_k = \begin{cases}
    1 &\quad \text{if } a_k = 1 \text{ and } S_k = 1,\\
    i &\quad \text{if } a_k = 1 \text{ and } S_k \in \{ia,ib\} \text{ for } i \in [K]\setminus \{1\},\\
    j &\quad \text{if } a_k = 0 \text{ and } S_k \in \{jc,jd\} \text{ for } j \in [L].
\end{cases}
\]
Hence, in the previous example, we have $a = (0,1,0,1,0,1,0)$ and $b = (1,2,1,1,2,2,2)$. Moreover, we define $\cI_k$ and $\cJ_k$ as the sets of indices for previously decoded channel inputs and quantization codewords respectively, i.e., for $k \in [2(K+L)-1]$, 
\begin{align*}
    \cI_k &= \{i \in \cP_1: S_\ell = i \text{ for some } \ell \in [k-1]\},\\
    \cJ_k &= \{i \in \cP_2: S_\ell = i \text{ for some } \ell \in [k-1]\}.\\[-2em]
\end{align*}
In the previous example, when $k=5$, we have $\cI_5 = \{1,2a\}$ and $\cJ_5 = \{1c,1d\}$. To the $k$th decoded virtual input/quantization codeword, we associate the rate
\[
\beta_{S_k} = \begin{cases}
    I(X_{S_k} ; \Yh_{\cJ_k}, X_{\cI_k}) &\quad \text{if } a_k =1,\\
    I(Y_{b_k}; \Yh_{S_k} \cond \Yh_{\cJ_k}, X_{\cI_k})&\quad \text{if } a_k =0,
\end{cases}
\]

\begin{figure*}[t]
\vspace{-2em}
	\centering
	\hspace*{5em}
 \small
	\def\svgscale{0.9}
	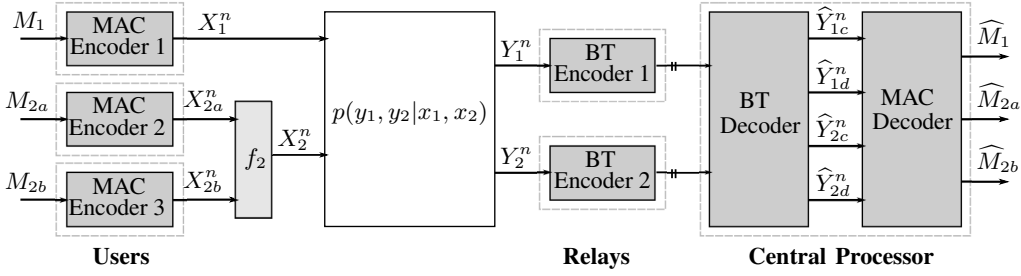
	\caption{Coding scheme for the two-user, two-relay uplink C-RAN problem using a multiple-access channel code and a Berger--Tung code.}
	\label{fig:uplink_cran_coding_scheme}
 \vspace{-1em}
\end{figure*} 

\noindent This defines a rate-fronthaul tuple $(R_1, \ldots, R_K, C_1, \ldots, C_L)$ such that $R_1 = \beta_1$, $R_k = \beta_{ka} + \beta_{kb}$ for each $k \in [K]\setminus \{1\}$, and $C_\ell = \beta_{\ell c} + \beta_{\ell d}$ for each $\ell \in [L]$. One can check that
\begin{align*}
    &C([L]) - R([K]) \\
    &=\textstyle{\sum_{\substack{k=1:\\a_k=0}}^{M}} I(Y_{b_k}; \Yh_{S_k} \cond \Yh_{\cJ_k}, X_{\cI_k}) - \textstyle{\sum_{\substack{k=1:\\a_k=1}}^{M}} I(X_{S_k} ; \Yh_{\cJ_k}, X_{\cI_k})\\
    &= \textstyle{\sum_{\substack{k=1:\\a_k=0}}^{M}} \left(I(Y_{b_k}; \Yh_{S_k}) - I(\Yh_{S_k}; \Yh_{\cJ_k})\right)  \\
    & \quad - \textstyle{\sum_{\substack{k=1:\\a_k=0}}^{M}} I(\Yh_{S_k}; X_{\cI_k}\cond \Yh_{\cJ_k}) - \textstyle{\sum_{\substack{k=1:\\a_k=1}}^{M}} I(X_{S_k} ; \Yh_{\cJ_k}, X_{\cI_k})\\
    &= I(Y_1^L; \Yh_1^L) - \textstyle{\sum_{\substack{k=1:\\a_k=0}}^{M}} H(\Yh_{S_k} \cond \Yh_{\cJ_k}) - \textstyle{\sum_{\substack{k=1:\\a_k=1}}^{M}} H(X_{S_k})  \\
    &\quad + \textstyle{\sum_{\substack{k=1:\\a_k=0}}^{M}} H(\Yh_{S_k} \cond X_{\cI_k},\Yh_{\cJ_k}) + \textstyle{\sum_{\substack{k=1:\\a_k=1}}^{M}} H(X_{S_k} \cond \Yh_{\cJ_k} , X_{\cI_k})\\
    &= I(Y_1^L;\Yh_1^L) - H(\Yh_1^L) - H(X_1^K) + H(X_1^K,\Yh_1^L) \\
    &= I(Y_1^L;\Yh_1^L) - I(X_1^K;\Yh_1^L)= I(Y_1^L;\Yh_1^L \cond X_1^K),
\end{align*}
where we used $M=2(K+L)-1$. Hence, $(R_1, \ldots, R_K,$ $C_1, \ldots, C_L) \in \cD$. Therefore, this defines a continuous function
$
\Psi: [0,1]^{K+L-1} \to \cD,
$
which maps the parameter vector $\alpha$ into a point on the dominant face $\cD$. The main ingredient of the proof of Theorem~\ref{thm:uplink} is to show that this mapping is onto, i.e., for each point on the dominant face $\cD$ of the joint decoding rate region, there exists a parameter vector $\alpha$ (and, hence, a splitting parameter vector $\eps$ and a decoding order $S$) that can achieve this point through splitting each channel input and quantization codeword at most once. To show that the mapping $\Psi$ is onto, one can use, as in~\cite{Grant2001}, the notion of \emph{degree} of a continuous function. Namely, if a continuous function that is defined over two convex polytopes has a nonzero degree, and if the function maps the boundary of its domain to the boundary of its range, then the function should be onto. To show that the sufficient conditions hold for our function $\Psi$, the proof would follow very similarly as that of~\cite[Theorem 10]{Grant2001}, and, hence, the details are omitted here. Note that~\cite{Grant2001} uses the structure of the dominant face of the multiple-access channel capacity region; therefore, one should, wherever needed, instead utilize the structure of the dominant face of the joint decoding rate region of uplink C-RAN's, which we have precisely established in Section~\ref{sec:dominant_uplink}.

\vspace{-.5em}
\subsection{Coding for Uplink C-RAN's Using Point-to-Point Codes} \label{sec:p2p_uplink_sub}

The result that the function $\Psi$ is continuous and onto implies that, by splitting each channel input and quantization codeword once, one can represent every point on the dominant face of the $K$-user, $L$-relay rate region as a \emph{corner point} in the higher-dimensional rate region of a $(2K-1)$-user, $(2L)$-relay uplink C-RAN problem with ``virtual'' inputs $(X_1,X_{2a},X_{2b},\ldots, X_{Ka}, X_{Kb})$ and ``virtual'' outputs $(\hat{Y}_{1c}, \Yh_{1d},\ldots, \Yh_{Lc},\Yh_{Ld})$. The proof of Theorem~\ref{thm:uplink} is then completed by constructing a coding scheme that can achieve the corner points of the higher-dimensional rate region using point-to-point codes. For this purpose, we refer to the Lego-brick approach to coding~\cite[Appendix L]{Ghaddar2024}, where an explicit code construction for the $K$-user, $L$-relay uplink C-RAN problem is given that uses at most $2(K+L)$ point-to-point codes designed for symmetric channels and achieves a corner point in the joint decoding rate region. Note that, by the equivalence of joint decoding and successive decoding (Theorem~\ref{thm:equivalence-uplink}), it holds that the construction in~\cite{Ghaddar2024} can achieve \emph{any} corner point in the joint decoding rate region of uplink C-RAN's. Hence, by using the same construction to code for the higher-dimensional $(2K-1)$-user, $(2L)$-relay uplink C-RAN problem, one needs at most $4(K+L)-2$ point-to-point codes that are designed for symmetric channels to achieve any point on the dominant face of the joint decoding rate region (not just the corner points). This completes the proof of Theorem~\ref{thm:uplink}.

Fig.~\ref{fig:uplink_cran_coding_scheme} shows the application of the code constructions in~\cite{Ghaddar2024} to the two-user, two-relay uplink C-RAN problem with rate splitting and quantization splitting. The coding scheme uses a multiple-access channel code and a Berger--Tung code. The multiple-access channel code is designed for the channel $p(\yh_{1c},\yh_{1d}, \yh_{2c},\yh_{2d} \cond x_1,x_{2a},x_{2b})$ and targets channel input distributions $p(x_1)p(x_{2a})p(x_{2b})$, which are derived from the rate splitting and quantization splitting procedures. The Berger--Tung code is designed for a source distribution $p(y_1,y_2)$ and targets source-reconstruction conditional distributions $p(\yh_{1c},\yh_{1d} \cond y_1)p(\yh_{2c},\yh_{2d} \cond y_2)$, which are due to quantization splitting. Note that the multiple-access channel code can be implemented using six point-to-point channel codes, and the Berger--Tung code can be implemented using eight point-to-point channel codes, as described in Appendices G and I of~\cite{Ghaddar2024}, respectively. Finally, we note that the coding scheme of Fig.~\ref{fig:uplink_cran_coding_scheme} successively decodes the quantization codewords first, and then the user messages; nonetheless, other decoding orders can be readily employed by modifying the design channels of the constituent point-to-point channel codes.

\section{The Downlink C-RAN Problem} \label{sec:downlink}

In this section, we consider the downlink C-RAN problem, and show analogous results as those presented in Sections~\ref{sec:equivalence_uplink} and~\ref{sec:p2p_uplink} for the uplink C-RAN problem. First, we show the equivalence of the rate regions that can be achieved by joint and successive encoding, as defined in Section~\ref{sec:preliminaries_downlink}.

\vspace{-1em}
\subsection{Equivalence of Joint Encoding sand Successive Encoding}
The main result of this part is the following theorem.

\begin{theorem} \label{thm:equivalence_downlink}
	The achievable rate region of joint encoding is equivalent to that of successive encoding for the downlink C-RAN problem, i.e.,
	$\cR_{JE} = \cR_{SE}.$
	
\end{theorem}

\noindent To prove Theorem~\ref{thm:equivalence_downlink}, we describe a procedure to find the corner points of the rate region $\cR_{JE}$. This can be done by finding the intersection points of the hyperplanes defined by the set of inequalities in~(\ref{eqn:rate_downlink_ddf}). As in the uplink setting, we will derive the coordinates of a corner point of $\cR_{JE}$ by solving for the intersection points in some prescribed order. Towards, this end, let $S^{K+L}$ be a permutation of $(R_1,\ldots, R_K, C_1,\ldots, C_L)$ that is set iteratively as in equation~(\ref{eqn:procedure_iterative_downlink}) at the top of this page, where $(\cI_k,\cJ_k,a_k,b_k)$ are defined as in~(\ref{eqn:IkJk}) and~(\ref{eqn:akbk}). Note that the iterative procedure corresponds to setting the inequality in~(\ref{eqn:rate_downlink_ddf}) to equality, with
\begin{equation} \label{eqn:set_choice_downlink}
	\begin{cases}
		\cT = \cI_k \cup \{b_k\} \text{ and } \cS = \cJ_k &\text{ if } a_k=1,\\
		\cT = \cI_k \text{ and } \cS = \cJ_k \cup \{b_k\} &\text{ if } a_k=0.
	\end{cases}
\end{equation}

\begin{figure*}
\vspace{-1.5em}
    \fbox{\parbox{\textwidth}{\textbf{Iterative procedure:} Given a permutation $S^{K+L}$ of $(R_1,\ldots, R_K, C_1,\ldots, C_L)$, for each $k\in [K+L]$, set
    \begin{equation} \label{eqn:procedure_iterative_downlink}
    	S_k \hspace*{-0.1em}=\hspace*{-0.1em} \begin{cases}
    		C(\cJ_k) \hspace*{-0.1em} \hspace*{-0.1em}-\hspace*{-0.1em} R(\cI_k)\hspace*{-0.1em} + \hspace*{-0.1em} \sum_{\substack{i \in \cI_k\cup \{b_k\}}}I(U_i;Y_i)  \hspace*{-0.1em} -\hspace*{-0.1em} I^{*}(U_{\cI_k\cup \{b_k\}}) \hspace*{-0.1em}-\hspace*{-0.1em} I^{*}(X_{\cJ_k}) \hspace*{-0.1em}-\hspace*{-0.1em} I(U_{\cI_k}, U_{b_k}; X_{\cJ_k}) &\hspace*{-0.3em}\text{ if } a_k = 1,\\
    		R(\cI_k)\hspace*{-0.1em}-\hspace*{-0.1em}C(\cJ_k) - \sum_{i \in \cI_k}\hspace*{-0.1em}I(U_i;Y_i) \hspace*{-0.1em}+\hspace*{-0.1em} I^{*}(U_{\cI_k}) \hspace*{-0.1em}+\hspace*{-0.1em} I^{*}(X_{\cJ_k\cup \{b_k\}}) \hspace*{-0.1em}+\hspace*{-0.1em} I(U_{\cI_k}; X_{\cJ_k}, X_{b_k}) &\hspace*{-0.3em}\text{ if } a_k = 0.
    	\end{cases}
    \end{equation}
    }}
    \vspace{-1.5em}
\end{figure*}

The following two lemmas show that the procedure described by~(\ref{eqn:procedure_iterative_downlink}) gives an explicit way of finding any corner point of the joint encoding rate region, which can be seen as an analogous result as that of Lemmas~\ref{lemma:procedure_iterative_modified} and~\ref{lemma:corner_point} for the uplink C-RAN case. We omit the details of the proofs as they follow very similarly as in Lemmas~\ref{lemma:procedure_iterative_modified} and~\ref{lemma:corner_point}.

\begin{lemma} \label{lemma:procedure_iterative_modified_downlink}
	The procedure described by equations~(\ref{eqn:procedure_iterative_downlink}) is equivalent to setting, for each $k \in [K+L]$,
	\begin{equation} \label{eqn:procedure_iterative_modified_downlink}
		S_k = \begin{cases}
			I(U_{b_k};Y_{b_k}) - I(U_{b_k};U_{\cI_k},X_{\cJ_k})& \text{ if } a_k = 1,\\
			I(X_{b_k}; U_{\cI_k},X_{\cJ_k})& \text{ if } a_k = 0.
		\end{cases}
	\end{equation}

\end{lemma}

\begin{lemma} \label{lemma:corner_point_downlink}
	The procedure described by~(\ref{eqn:procedure_iterative_downlink}) gives a corner point of the rate region $\cR_{DDF}$.
\end{lemma}

The proof of Theorem~\ref{thm:equivalence_downlink} then follows by showing that every corner point of $\cR_{JE}$ belongs to $\cR_{SE}$, where a corner point of $\cR_{JE}$ is completely defined in~(\ref{eqn:procedure_iterative_modified_downlink}) for some permutation $S^{K+L}$ of $(R_1,\ldots, R_K, C_1,\ldots, C_L)$. Hence, by defining the encoding order $\pi = (\pi_1,\ldots,\pi_{K+L})$ as
\begin{equation}
	\pi_k = \begin{cases}
		U_{b_{k}} &\quad \text{ if } a_k = 1,\\
		X_{b_{k}} &\quad \text{ if } a_k = 0,
	\end{cases}
\end{equation}
and comparing the expression~(\ref{eqn:procedure_iterative_modified_downlink})  with the rate region~(\ref{eqn:rate_downlink_gcs}) of successive encoding, it follows that the corner point $S^{K+L}$ of $\cR_{JE}$ is also a corner point of $\cR_{SE}$ when the decoding order $\pi$ is used. This completes the proof of Theorem~\ref{thm:equivalence_downlink}.

% \noindent \emph{Proof of Theorem~\ref{thm:equivalence_downlink}.} Now, we are ready to prove Theorem~\ref{thm:equivalence_downlink}. Clearly, we have that $\cR_{SE} \subseteq \cR_{JE}$. To show the other direction, it suffices to show that every corner point of $\cR_{JE}$ belongs to $\cR_{SE}$. From Lemmas~\ref{lemma:procedure_iterative_modified_downlink} and~\ref{lemma:corner_point_downlink}, we know that a corner point of $\cR_{JE}$ can be expressed iteratively as follows:
% \begin{equation} \label{eqn:procedure_modified_downlink_2}
% 	S_k = \begin{cases}
% 		I(U_{b_k};Y_{b_k}) - I(U_{b_k};U_{\cI_k},X_{\cJ_k})&\quad \text{ if } a_k = 1,\\
% 		I(X_{b_k}; U_{\cI_k},X_{\cJ_k})&\quad \text{ if } a_k = 0.
% 	\end{cases}
% \end{equation}
% for each $k \in [K+L]$, where $S^{K+L}$ is some permutation of $(R_1,\ldots, R_K, C_1,\ldots, C_L)$, and $(\cI_k,\cJ_k, a_k,b_k)$ are defined as in~(\ref{eqn:IkJk}) and~(\ref{eqn:akbk}). Let $\pi = (\pi_1,\ldots,\pi_{K+L})$ be a permutation of $(U_1,\ldots,U_K, X_1,\ldots, X_L)$ such that
% \begin{equation}
% 	\pi_k = \begin{cases}
% 		U_{b_{k}} &\quad \text{ if } a_k = 1,\\
% 		X_{b_{k}} &\quad \text{ if } a_k = 0,
% 	\end{cases}
% \end{equation}
% for each $k \in [K+L]$. Comparing the expression~(\ref{eqn:procedure_modified_downlink_2}) with the rate region~(\ref{eqn:rate_downlink_gcs}) of the generalized compression strategy, it can be seen that the corner point $S^{K+L}$ of $\cR_{JE}$ is also a corner point of $\cR_{SE}$ when the decoding order $\pi$ is used. This completes the proof.

\vspace{-1em}
\subsection{Achieving Joint Encoding Region via Point-to-Point Codes}
Similar to the uplink setting, one can construct a coding scheme that achieves any rate-fronthaul point in the joint encoding rate region for the downlink C-RAN using at most $4(K+L)-2$ point-to-point codes. For space limitations, and since the argument follows very similarly to existing work and to derivations that have been made in this paper so far, the details of the proof will be omitted.

The construction draws inspiration from~\cite{Lele2021}, where a splitting technique (that is similar to quantization splitting) is used to achieve any point in the rate region of Marton coding for the $K$-user broadcast channel with at most $2K-1$ splits of the desired input distribution. A similar splitting approach can be followed for the target distribution $p(u_1,\ldots,u_K,x_1,\ldots,x_L)$ of the downlink C-RAN problem. As in~\cite{Lele2021}, a similar inductive argument can be followed to show that at most $2(K+L)-1$ splits of the target distribution are needed to achieve any point on the joint encoding rate region. Using the code construction of downlink C-RAN's from~\cite[Appendix K]{Ghaddar2024}, it follows that one can construct a coding scheme that achieves any rate point in the joint encoding rate region using at most $4(K+L)-2$ point-to-point codes.

\section{Concluding Remarks} \label{sec:conclusion}
This paper presents a technique for achieving the entire rate-fronthaul region of joint coding and compression for uplink and downlink C-RAN's. The technique presented is based on rate splitting and quantization splitting, which are developed in the settings of coding for multiple-access channels and distributed lossy compression, respectively. By leveraging the code constructions in the recently-developed Lego-brick framework~\cite{Ghaddar2024}, we are able to construct coding schemes that achieve the entire rate-fronthaul region using point-to-point channel codes that are designed for symmetric channels. Note that the computational complexity of the proposed code construction is governed by the encoding/decoding complexities of the constituent point-to-point codes, which can be taken ``off the shelf''; hence, the coding schemes are friendly to hardware implementation using existing coding blocks. As future work, it is of particular interest to generalize the constructions in~\cite{Ghaddar2024} to utilize \emph{binary} point-to-point channel codes when coding for channels with \emph{non-binary} inputs. This would allow one to implement coded modulation schemes for general multi-user settings, including cloud radio access networks.

\begin{appendices}
    \renewcommand{\thesectiondis}[2]{\Alph{section}:}

\vspace{-.5em}
\section{Proof of Lemma~\ref{lemma:procedure_iterative_modified}} \label{appx:procedure_iterative_modified}
    When $k=1$, we have $\cI_1 = \cJ_1 = \emptyset$. The result clearly holds when $a_1 = 1$. When $a_1 = 0$, we have
    \begin{align*}
        &\quad I(Y_{b_1}; \hat{Y}_{b_1}) - I(\hat{Y}_{b_1}; X_1, \ldots, K, \hat{Y}_{[L] \setminus \{b_1\}}) \\[-.5em]
        &\stackrel{(a)}{=} H(\hat{Y}_{b_1} \cond X_1, \ldots, X_K) - H(\hat{Y}_{b_1} \cond Y_{b_1})\\[-.5em]
        &\stackrel{(b)}{=} I(Y_{b_1};\hat{Y}_{b_1} \cond X_1,\ldots, X_K),
    \end{align*}
    where $(a)$ follows since $\hat{Y}_{b_1}$ is independent of $\hat{Y}_{[L]\setminus \{b_1\}}$ given $(X_1,\ldots, X_K)$, and $(b)$ follows since $\hat{Y}_{b_1}$ is independent of $(X_1,\ldots, X_K)$ given $Y_{b_1}$.
    
    For $k>1$, we proceed by induction. Let us first consider the case when $a_{k-1} = 1$ and $a_k = 1$. Thus, we have $\cI_k = \cI_{k-1} \cup \{b_{k-1}\}$, $\cJ_k = \cJ_{k-1}$ and $S_{k-1} = R_{b_{k-1}}$. Hence, by the iterative procedure in~(\ref{eqn:procedure_iterative}), we have
    \begin{align*}
        S_k &= C(\cJ_{k-1}) - R(\cI_{k-1}) - R_{b_{k-1}}  - I(Y_{\cJ_{k-1}};\hat{Y}_{\cJ_{k-1}} \cond X_1^K) \\
        &\quad + I(X_{b_k}, X_{\cI_k}; \hat{Y}_{\cJ_k^c} \cond X_{\cI_k^c\setminus \{b_k\}})\\[-.5em]
        &\stackrel{(c)}{=} - I(X_{b_{k-1}}, X_{\cI_{k-1}}; \hat{Y}_{\cJ_{k-1}^c} \cond X_{\cI_{k-1}^c\setminus \{b_{k-1}\}}) \\
        &\quad + I(X_{b_k}, X_{\cI_k}; \hat{Y}_{\cJ_k^c} \cond X_{\cI_k^c\setminus \{b_k\}})\\
        &= -I(X_{\cI_k};\hat{Y}_{\cJ_k^c}\cond X_{\cI_k^c})+ I(X_{b_k}, X_{\cI_k}; \hat{Y}_{\cJ_k^c} \cond X_{\cI_k^c\setminus \{b_k\}})\\
        &= I(X_{b_k}; \hat{Y}_{\cJ_k^c} \cond X_{\cI_k^c \setminus \{b_k\}}),\\[-2em]
    \end{align*}
    where in $(c)$ we used the induction hypothesis. 
    % Now, suppose $a_{k-1} = 0$ and $a_k = 1$. In this case, $\cI_k = \cI_{k-1}$, $\cJ_k = \cJ_{k-1} \cup \{b_{k-1}\}$ and $S_{k-1} = C_{b_{k-1}}$. By~(\ref{eqn:procedure_iterative}), we have
    % \begin{align*}
    %     S_k &= C(\cJ_{k-1}) - I(Y_{\cJ_{k-1} \cup \{b_{k-1}\}};\hat{Y}_{\cJ_{k-1}\cup \{b_{k-1}\}} \cond X_1^K) \\
    %     &\quad + C_{b_{k-1}} -R(\cI_{k-1})  + I(X_{b_k}, X_{\cI_k}; \hat{Y}_{\cJ_k^c} \cond X_{\cI_k^c\setminus \{b_k\}})\\
    %     &\stackrel{(d)}{=} - I(X_{\cI_{k-1}};\hat{Y}_{\cJ_{k-1}^c\setminus \{b_{k-1}\}} \cond X_{\cI_{k-1}^c}) \\
    %     &\quad + I(X_{b_k}, X_{\cI_k}; \hat{Y}_{\cJ_k^c} \cond X_{\cI_k^c\setminus \{b_k\}})\\
    %     &= - I(X_{\cI_{k}};\hat{Y}_{\cJ_k^c} \cond X_{\cI_k^c}) + I(X_{b_k}, X_{\cI_k}; \hat{Y}_{\cJ_k^c} \cond X_{\cI_k^c\setminus \{b_k\}})\\
    %     &= I(X_{b_k}; \hat{Y}_{\cJ_k^c} \cond X_{\cI_k^c \setminus \{b_k\}}),
    % \end{align*}
    % where $(d)$ uses the induction hypothesis. 
    Next, suppose that $a_{k-1} = 1$ and $a_k = 0$. Thus, we have $\cI_k = \cI_{k-1} \cup \{b_{k-1}\}$, $\cJ_k = \cJ_{k-1}$ and $S_{k-1} = R_{b_{k-1}}$. Hence, by the iterative procedure in~(\ref{eqn:procedure_iterative}), we have
    \begin{align*}
        &S_k =R(\cI_{k-1}) + R_{b_{k-1}} - C(\cJ_{k-1}) - I(X_{\cI_k};\hat{Y}_{\cJ_k^c\setminus \{b_k\}} \cond X_{\cI_k^c})\\
        &\quad+ I(Y_{\cJ_{k-1} \cup \{b_k\}};\hat{Y}_{\cJ_{k-1} \cup \{b_k\}} \cond X_1^K) \\
        &=R(\cI_{k-1}) + R_{b_{k-1}} - C(\cJ_{k-1}) - I(X_{\cI_k};\hat{Y}_{\cJ_k^c\setminus \{b_k\}} \cond X_{\cI_k^c}) \\
        &\quad + I(Y_{\cJ_{k-1}};\hat{Y}_{\cJ_{k-1}} \cond X_1^K) + I(Y_{b_k};\hat{Y}_{b_k} \cond X_1^K)\\
        &= I(X_{b_{k-1}}, X_{\cI_{k-1}}; \hat{Y}_{\cJ_{k-1}^c} \cond X_{\cI_{k-1}^c\setminus \{b_{k-1}\}}) + I(Y_{b_k};\hat{Y}_{b_k}\cond X_1^K) \\
        &\quad- I(X_{\cI_k};\hat{Y}_{\cJ_k^c\setminus \{b_k\}} \cond X_{\cI_k^c})\\
        %&= I(X_{\cI_k}; \hat{Y}_{\cJ_k^c} \cond X_{\cI_k^c}) + I(Y_{b_k};\hat{Y}_{b_k}\cond X_1^K) - I(X_{\cI_k};\hat{Y}_{\cJ_k^c\setminus \{b_k\}} \cond X_{\cI_k^c})\\
        &= I(Y_{b_k};\hat{Y}_{b_k}\cond X_1^K) + I(X_{\cI_k}; \hat{Y}_{b_k} \cond X_{\cI_k^c}, \hat{Y}_{\cJ_k^c\setminus \{b_k\}})\\[-.5em]
        &\stackrel{(e)}{=} H(\hat{Y}_{b_k} \cond X_{\cI_k^c}, \hat{Y}_{\cJ_k^c\setminus \{b_k\}}) - H(\hat{Y}_{b_k} \cond Y_{b_k})\\
        &= I(Y_{b_k}; \hat{Y}_{b_k}) - I(\hat{Y}_{b_k}; X_{\cI_k^c}, \hat{Y}_{\cJ_k^c \setminus \{b_k\}}),\\[-2em]
    \end{align*}
    where $(e)$ follows since $\hat{Y}_{b_k}$ is independent of $\hat{Y}_{\cJ_k^c\setminus \{b_k\}}$ given $(X_1,\ldots, X_K)$, and $\hat{Y}_{b_k}$ is independent of $(X_1,\ldots, X_K)$ given $Y_{b_k}$. The result for the remaining cases of $(a_{k-1}, a_k)$ follows similarly as the previous two cases, and is omitted for brevity.

\vspace{-.5em}
\section{Proof of Lemma~\ref{lemma:corner_point}} \label{appx:corner_point}
    Let us first consider the case of $a_k = 1$. Recall that the procedure described by~(\ref{eqn:procedure_iterative}) corresponds to setting the inequality in~(\ref{eqn:rate_uplink_joint}) to equality, with
$
        \cS = \cS_1 \triangleq \cI_k \cup \{b_k\} \text{ and } \cT = \cT_1 \triangleq \cJ_k
$
    We want to show that the inequality in~(\ref{eqn:rate_uplink_joint}) would be inactive for any other choice of $\cS$ and $\cT$ that allows to solve for $S_k$. More precisely, we want to show that by setting the inequality in~(\ref{eqn:rate_uplink_joint}) to equality with $\cS = \cS_2 \triangleq \cA \cup \{b_k\}$ and $\cT =\cT_2 \triangleq  \cB$ for any arbitrary sets $\cA \subseteq \cI_k$ and $\cB \subseteq \cJ_k$, we get a larger communication rate for $R_{b_k}$, and hence, the corresponding inequality is inactive. Note that for our choice of $\cS=\cS_1$ and $\cT=\cT_1$, we have by Lemma~\ref{lemma:procedure_iterative_modified} that $R_{b_k} = S_k = I(X_{b_k}; \hat{Y}_{\cJ_k^c} \cond X_{\cI_k^c \setminus \{b_k\}})$, while under the choice of $\cS=\cS_2$ and $\cT=\cT_2$, we get
    \begin{align*}
        \tilde{R}_{b_k} &= C(\cB) - R(\cA) - I(Y_\cB;\hat{Y}_\cB \cond X_1,\ldots, X_K) \\
        &\quad + I(X_{b_k}, X_{\cA}; \hat{Y}_{\cB^c} \cond X_{\cA^c\setminus \{b_k\}})\\[-.5em]
        &\stackrel{(a)}{\geq} - I(X_\cA;\hat{Y}_{\cB^c} \cond X_{\cA^c}) + I(X_{b_k}, X_{\cA}; \hat{Y}_{\cB^c} \cond X_{\cA^c\setminus \{b_k\}})\\
        &= I(X_{b_k}; \hat{Y}_{\cB^c} \cond X_{\cA^c\setminus \{b_k\}}),
    \end{align*}
    where $(a)$ holds by the inequalities governed by the rate region~(\ref{eqn:rate_uplink_joint}) with $\cS = \cA$ and $\cT = \cB$. It follows that 
    \begin{align*}
        &\tilde{R}_{b_k} -R_{b_k} \geq I(X_{b_k}; \hat{Y}_{\cB^c} \cond X_{\cA^c\setminus \{b_k\}})- I(X_{b_k}; \hat{Y}_{\cJ_k^c} \cond X_{\cI_k^c \setminus \{b_k\}})\\[-.5em]
        &\stackrel{(b)}{=}I(X_{b_k}; \hat{Y}_{\cB^c}, X_{\cA^c\setminus \{b_k\}})- I(X_{b_k}; \hat{Y}_{\cJ_k^c}, X_{\cI_k^c \setminus \{b_k\}})
        \stackrel{(c)}{\geq} 0,
    \end{align*}
    where $(b)$ follows since $(X_1,\ldots, X_K)$ are mutually independent, and $(c)$ follows since $\cA \subseteq \cI_k$ and $\cB \subseteq \cJ_k$.\\[0.5em]
	Now, consider the case of $a_k=0$. The procedure described by~(\ref{eqn:procedure_iterative}) corresponds to setting the inequality in~(\ref{eqn:rate_uplink_joint}) to equality, with
$
        \cS = \cS_3 \triangleq \cI_k \text{ and } \cT = \cT_3 \triangleq \cJ_k \cup \{b_k\}.
$
    We want to show that by setting the inequality in~(\ref{eqn:rate_uplink_joint}) to equality with $\cS = \cS_4 \triangleq \cA$ and $\cT =\cT_4 \triangleq  \cB \cup \{b_k\}$, we get a smaller compression rate for $C_{b_k}$. Note that for our choice of $\cS=\cS_3$ and $\cT=\cT_3$, we have by Lemma~\ref{lemma:procedure_iterative_modified} that
    $
    C_{b_k} = S_k = I(Y_{b_k}; \hat{Y}_{b_k}) - I(\hat{Y}_{b_k}; X_{\cI_k^c}, \hat{Y}_{\cJ_k^c \setminus \{b_k\}}),
    $
    while under the choice $\cS = \cS_4$ and $\cT = \cT_4$, we get
    \begin{align*}
         \tilde{C}_{b_k} &= R(\cA) - C(\cB) + I(Y_{\cB}, Y_{b_k};\hat{Y}_\cB, \Yh_{b_k} \cond X_1^K) \\
         &\quad - I(X_{\cA};\hat{Y}_{\cB^c\setminus \{b_k\}} \cond X_{\cA^c})\\
         &= R(\cA) - C(\cB) + I(Y_{\cB};\hat{Y}_\cB \cond X_1^K) + I(Y_{b_k};\Yh_{b_k} \cond X_1^K)\\
         &\quad - I(X_{\cA};\hat{Y}_{\cB^c\setminus \{b_k\}} \cond X_{\cA^c})\\[-.5em]
         &\stackrel{(d)}{\leq} I(Y_{b_k};\Yh_{b_k} \cond X_1^K) + I(X_{\cA};\hat{Y}_{\cB^c} \cond X_{\cA^c}) \\
         &\quad - I(X_{\cA};\hat{Y}_{\cB^c\setminus \{b_k\}} \cond X_{\cA^c})\\[-.5em]
         &\stackrel{(e)}{=} I(Y_{b_k};\Yh_{b_k}) - I( \Yh_{b_k}; X_1^K) + I(X_{\cA};\hat{Y}_{b_k} \cond X_{\cA^c}, \hat{Y}_{\cB^c\setminus \{b_k\}})\\[-.5em]
         &\stackrel{(f)}{=}  I(Y_{b_k};\Yh_{b_k}) - I(\Yh_{b_k}; X_{\cA^c}, \hat{Y}_{\cB^c\setminus \{b_k\}}),
    \end{align*}
    where $(d)$ holds by the inequalities governed by the rate region~(\ref{eqn:rate_uplink_joint}) with $\cS = \cA$ and $\cT = \cB$, $(e)$ follows since $\hat{Y}_{b_k}$ is independent of $(X_1,\ldots, X_K)$ given $Y_{b_k}$, and $(f)$ holds since $\hat{Y}_{b_k}$ is independent of $\hat{Y}_{\cB^c\setminus \{b_k\}}$ given $(X_1,\ldots, X_K)$ for any subset $\cB$. It follows that
    \begin{equation*}
        \tilde{C}_{b_k} - C_{b_k} \hspace*{-0.2em} \leq \hspace*{-0.2em} I(\hat{Y}_{b_k}; X_{\cI_k^c}, \hat{Y}_{\cJ_k^c \setminus \{b_k\}}) -  I(\Yh_{b_k}; X_{\cA^c}, \hat{Y}_{\cB^c\setminus \{b_k\}}) \hspace*{-0.2em} \leq \hspace*{-0.2em} 0,
    \end{equation*}
    where the last inequality holds since $\cA \subseteq \cI_k$ and $\cB \subseteq \cJ_k$. Therefore, in both cases $a_k = 1$ and $a_k = 0$, the only active inequality is the one corresponding to the choice of $\cS$ and $\cT$ given in~(\ref{eqn:set_choice}); hence, one can only consider this inequality when finding the coordinate $S_k$ of a corner point of $\cR_{JD}$, which is precisely the procedure done in~(\ref{eqn:procedure_iterative}).

\vspace{-.5em}
\section{Proof of Lemma~\ref{lemma:dominant}} \label{appx:dominant}
Define
$\cC = \big\{(R_1, \ldots, R_K, C_1, \ldots, C_L):\forall\, \cS \subseteq [K], \cT \subseteq [L],  I(Y_\cT;\Yh_\cT \cond X_1^K) - I(X_\cS;\Yh_{\cT^c} \cond X_{\cS^c}) \leq C(\cT)-R(\cS) \leq I(Y_\cT;\Yh_\cT \cond X_\cS)\big\}.$
% \begin{align*}
%     \cC &= \Big\{(R_1, \ldots, R_K, C_1, \ldots, C_L): \forall \, \cS \subseteq [K], \cT \subseteq [L], \\
%     &\qquad I(Y_\cT;\Yh_\cT \cond X_1^K) - I(X_\cS;\Yh_{\cT^c} \cond X_{\cS^c}) \leq C(\cT)-R(\cS)\\
%     &\qquad \leq I(Y_\cT;\Yh_\cT \cond X_\cS)\Big\}.
% \end{align*}
Clearly, if $(R_1,\ldots,R_K,C_1,\ldots,C_L) \in \cC$, then $(R_1,\ldots,R_K,$ $C_1,\ldots,C_L) \in \cR_{JD}$, 
$
C([L]) - R([K]) = I(Y_1^L;\Yh_1^L \cond X_1^K),
$
and, thus, $(R_1,\ldots,R_K,C_1,\ldots,C_L) \in \cD$, and $\cC \subseteq \cD$. 

Conversely, let $(R_1,\ldots,R_K,C_1,\ldots,C_L) \in \cD$, and consider $\cS \subseteq [K]$ and $\cT \subseteq [L]$. Since $(R_1,\ldots,R_K,$ $C_1,\ldots,C_L) \in \cR_{JD}$, then
$
C(\cT)-R(\cS) \geq I(Y_\cT;\Yh_\cT \cond X_1^K) - I(X_\cS;\Yh_{\cT^c} \cond X_{\cS^c}).
$
Moreover,
\begin{align*}
    &C(\cT) -R(\cS) = C([L]) - R([K]) - \left(C(\cT^c) - R(\cS^c)\right)\\[-.5em]
    &\stackrel{(a)}{\leq} I(Y_1^L;\Yh_1^L \cond X_1^K) \hspace*{-0.1em} -  \hspace*{-0.1em} \left(I(Y_{\cT^c}; \Yh_{\cT^c} \cond X_1^K) \hspace*{-0.1em} - \hspace*{-0.1em} I(X_{\cS^c};\Yh_\cT \cond X_\cS)\right)\\
    &= I(Y_{\cT}; \Yh_{\cT} \cond X_1,\ldots,X_K)+ I(X_{\cS^c};\Yh_\cT \cond X_\cS)\\
    &= I(Y_{\cT}; \Yh_{\cT}) - I(\Yh_{\cT}; X_1,\ldots,X_K)+ I(X_{\cS^c};\Yh_\cT \cond X_\cS)\\
    &= I(Y_{\cT}; \Yh_{\cT}) - I(\Yh_{\cT}; X_\cS)=I(Y_{\cT}; \Yh_{\cT} \cond X_\cS),\\[-2em]
\end{align*}
where $(a)$ holds since $(R_1,\ldots,R_K,C_1,\ldots,C_L) \in \cD$. Therefore, $\cD \subseteq \cC$, and the result follows.

\vspace{-.5em}
\section{Proof of Lemma~\ref{lemma:boundary}} \label{appx:boundary}
Let $(R_1,\ldots,R_K,C_1,\ldots,C_L) \in \cF_{\cS,\cT}$, i.e., $(R_1,\ldots,R_K,$ $C_1,\ldots,C_L) \in \cD$ and $C(\cT) - R(\cS) = I(Y_{\cT};\Yh_\cT \cond X_\cS)$. Let $\cA \subseteq \cS,\cB \subseteq \cT.$ Then $C(\cB) - R(\cA) \leq I(Y_{\cB};\Yh_\cB \cond X_\cA)$ and
\begin{align*}
    &C(\cB) - R(\cA) = C(\cT) - R(\cS) - (C(\cT\setminus\cB) - R(\cS \setminus\cA))\\
    &\geq I(Y_{\cT};\Yh_\cT \cond X_\cS) - I(Y_{\cT\setminus\cB}; \Yh_{\cT\setminus\cB} \cond X_{\cS\setminus\cA}) \\
    &= I(Y_\cB;\Yh_\cB) - I(\Yh_\cB;\Yh_{\cT\setminus\cB}) - I(X_\cS;\Yh_\cT) + I(X_{\cS\setminus\cA};\Yh_{\cT\setminus\cB})\\
    &=I(Y_\cB;\Yh_\cB) -I(\Yh_\cB;X_\cS, \Yh_{\cT\setminus\cB}) + I(\Yh_\cB;X_\cS \cond \Yh_{\cT\setminus\cB})\\
    &\quad - I(X_\cS;\Yh_\cT) + I(X_{\cS\setminus\cA};\Yh_{\cT\setminus\cB})\\
    &= I(Y_\cB;\Yh_\cB \cond X_\cS, \Yh_{\cT\setminus\cB}) - I(X_\cS;\Yh_{\cT\setminus\cB})+ I(X_{\cS\setminus\cA};\Yh_{\cT\setminus\cB})\\
    &=I(Y_\cB;\Yh_\cB \cond X_\cS, \Yh_{\cT\setminus\cB}) - I(X_\cA;\Yh_{\cT\setminus\cB} \cond X_{\cS\setminus\cA}).\\[-2em]
\end{align*}
Hence, $(R_\cS,C_\cT) \in \cD_{\cS,\cT}$. 
Let $\cA \subseteq \cS^c$ and $\cB \subseteq \cT^c$. We have
\begin{align*}
    &C(\cB) - R(\cA) \\
    &=C([L]) \hspace*{-0.1em} - \hspace*{-0.1em} R([K]) \hspace*{-0.1em} - \hspace*{-0.1em} C(\cT) \hspace*{-0.1em} + \hspace*{-0.1em} R(\cS) \hspace*{-0.1em} - \hspace*{-0.1em} C(\cT^c \hspace*{-0.1em} \setminus\hspace*{-0.1em}  \cB) \hspace*{-0.1em} + \hspace*{-0.1em} R(\cS^c \hspace*{-0.1em} \setminus \hspace*{-0.1em} \cA) \\
    &\leq I(Y_1^L;\Yh_1^L \cond X_1^K) - I(Y_\cT;\Yh_\cT \cond X_\cS) \\
    &\quad - \left( I(Y_{\cT^c \setminus \cB};\Yh_{\cT^c \setminus \cB} \cond X_1^K) - I(X_{\cS^c \setminus \cA};\Yh_\cT, \Yh_\cB \cond X_\cA,X_\cS)\right)\\
    &= I(Y_\cT;\Yh_\cT \cond X_1^K) + I(Y_\cB;\Yh_\cB \cond X_1^K) - I(Y_\cT;\Yh_\cT \cond X_\cS) \\
    &\quad + I(X_{\cS^c \setminus \cA};\Yh_\cT, \Yh_\cB \cond X_\cA,X_\cS) \\
    &= I(Y_\cB;\Yh_\cB) - I(X_1^K; \Yh_\cB) - I(X_{\cS^c}; \Yh_\cT \cond X_\cS)\\
    &\quad + I(X_{\cS^c \setminus \cA};\Yh_\cT, \Yh_\cB \cond X_\cA,X_\cS) \\
    &= I(Y_\cB;\Yh_\cB) - I(X_\cA, X_\cS,\Yh_\cT; \Yh_\cB) - I(X_\cA;\Yh_\cT,X_\cS)\\
    &= I(Y_\cB;\Yh_\cB\cond X_\cA, X_\cS,\Yh_\cT) - I(X_\cA;\Yh_\cT,X_\cS)\\
    &= f_{\cS^c,\cT^c \cond \cS,\cT}^{\mathrm{ub}}(\cA,\cB).\\[-1.8em]
\end{align*}
Similarly, one can show that $C(\cB) - R(\cA) \geq I(Y_1^L;\Yh_1^L \cond X_1^K) - I(Y_\cT;\Yh_\cT \cond X_\cS) - I(Y_{\cT^c\setminus\cB};\Yh_{\cT^c\setminus\cB} \cond X_{\cS^c\setminus\cA})$ $=f_{\cS^c,\cT^c \cond \cS,\cT}^{\mathrm{lb}}(\cA,\cB),$
% \begin{align*}
%     &C(\cB) - R(\cA) \geq I(Y_1^L;\Yh_1^L \cond X_1^K) - I(Y_\cT;\Yh_\cT \cond X_\cS)\\
%     &\quad - I(Y_{\cT^c\setminus\cB};\Yh_{\cT^c\setminus\cB} \cond X_{\cS^c\setminus\cA})=f_{\cS^c,\cT^c \cond \cS,\cT}^{\mathrm{lb}}(\cA,\cB),
% \end{align*}
and, hence, $(R_{\cS^c}, C_{\cT^c}) \in \cD_{\cS^c,\cT^c \cond \cS,\cT}$.

Conversely, let $(R_1,\ldots,R_K,C_1,\ldots, C_L)$ be such that $(R_\cS,C_\cT) \in \cD_{\cS,\cT}$ and $(R_{\cS^c}, C_{\cT^c}) \in \cD_{\cS^c,\cT^c \cond \cS,\cT}$. Then, we have $C(\cT)-R(\cS) = I(Y_\cT;\Yh_\cT \cond X_\cS)$, and
$
C(\cT^c)-R(\cS^c) = I(Y_{\cT^c};\Yh_{\cT^c}\cond X_{\cS^c},X_\cS,\Yh_\cT) - I(X_{\cS^c};\Yh_\cT,X_\cS).
$
% It follows that
% \begin{align*}
%     &C([L]) - R([K])  = C(\cT) + C(\cT^c) -R(\cS) -R(\cS^c)\\
%     &= I(Y_\cT;\Yh_\cT \cond X_\cS) + I(Y_{\cT^c};\Yh_{\cT^c}\cond X_{\cS^c},X_\cS,\Yh_\cT) - I(X_{\cS^c};\Yh_\cT,X_\cS)\\
%     &= I(Y_1^L;\Yh_1^L) - I(X_\cS; \Yh_\cT) - I(X_\cS; \Yh_{\cT^c} \cond \Yh_\cT)- I(X_{\cS^c};\Yh_{\cT^c},\Yh_\cT,X_\cS)\\
%     &= I(Y_1^L;\Yh_1^L) - I(X_1^K;\Yh_1^L)\\
%     &=I(Y_1^L;\Yh_1^L \cond X_1^K).
% \end{align*}
It follows that
\begin{align*}
    &C([L]) - R([K])  = C(\cT) + C(\cT^c) -R(\cS) -R(\cS^c)\\
    &=I(Y_\cT; \Yh_\cT) - I(X_\cS ; \Yh_\cT) + I(Y_{\cT^c}; \Yh_{\cT^c}) \\
    &\quad - I(\Yh_\cT; \Yh_{\cT^c}) - I(X_1^K; \Yh_{\cT^c} \cond \Yh_\cT) - I(X_{\cS^c};\Yh_\cT \cond X_\cS)\\
    % &= I(Y_\cT;\Yh_\cT \cond X_\cS) + I(Y_{\cT^c};\Yh_{\cT^c}\cond X_{\cS^c},X_\cS,\Yh_\cT) - I(X_{\cS^c};\Yh_\cT,X_\cS)\\
    % &= I(Y_1^L;\Yh_1^L) - I(X_\cS; \Yh_\cT) - I(X_\cS; \Yh_{\cT^c} \cond \Yh_\cT)- I(X_{\cS^c};\Yh_{\cT^c},\Yh_\cT,X_\cS)\\
    &= I(Y_1^L;\Yh_1^L) - I(X_1^K;\Yh_1^L)=I(Y_1^L;\Yh_1^L \cond X_1^K).\\[-2em]
\end{align*}
It remains to show that $C(\cB)-R(\cA) \leq I(Y_\cB;\Yh_\cB \cond X_\cA)$ for each $\cA \subseteq [K]$ and $\cB \subseteq [L]$. Let $\cA = \cM \cup \cN$ and $\cB = \cP \cup \cQ$, where $\cM \subseteq \cS$, $\cN \subseteq \cS^c$, $\cP \subseteq \cT$ and $\cQ \subseteq \cT^c$. Then,
\begin{align*}
    &C(\cB)  - R(\cA) = C(\cP) - R(\cM) + C(\cQ) - R(\cN)\\
    &\leq \! I(Y_\cP;\Yh_\cP \cond X_\cM) \! +\! I(Y_\cQ;\Yh_\cQ \cond \Yh_\cT,X_\cS)\! -\! I(X_\cN;\Yh_\cQ,\Yh_\cT,X_\cS)\\
    % &\quad - I(X_\cN;\Yh_\cQ,\Yh_\cT,X_\cS)\\
    &= I(Y_{\cP\cup\cQ}; \Yh_{\cP\cup \cQ}) + I(\Yh_\cP;\Yh_\cQ) - I(X_\cM;\Yh_\cP) \\
    &\quad - I(\Yh_\cQ;\Yh_\cT,X_\cS)- I(X_\cN;\Yh_\cQ,\Yh_\cT,X_\cS)\\[-.5em]
    &\stackrel{(a)}{\leq} I(Y_{\cP\cup\cQ}; \Yh_{\cP\cup \cQ}) + I(\Yh_\cP;\Yh_\cQ) - I(X_\cM;\Yh_\cP) \\
    &\quad - I(\Yh_\cQ;\Yh_\cP,X_\cM)- I(X_\cN;\Yh_\cQ,\Yh_\cP,X_\cM)\\
    &= I(Y_{\cP\cup\cQ}; \Yh_{\cP\cup \cQ}) - I(X_\cM;\Yh_\cP,\Yh_\cQ) - I(X_\cN;\Yh_\cQ,\Yh_\cP,X_\cM)\\
    &= I(Y_{\cP\cup\cQ}; \Yh_{\cP\cup \cQ}) - I(X_{\cM\cup\cN};\Yh_{\cP\cup\cQ})= I(Y_\cB;\Yh_\cB \cond X_\cA),\\[-2em]
\end{align*}
where $(a)$ follows since $\cM \subseteq \cS$ and $\cP \subseteq \cT$. Hence, $(R_1,\ldots,R_K,C_1,\ldots, C_L) \in \cF_{\cS,\cT}$, which proves the claim.

\vspace{-.5em}
\section{Proof of Lemma~\ref{lemma:degenerate-1}} \label{appx:degenerate-1}
Let us first start by showing that $(b)$ implies $(a)$. Let $(R_1,\ldots,R_K,C_1,\ldots,C_L) \in \cD$, i.e., $C([L]) - R([K]) = I(Y_1^L;\Yh_1^L\cond X_1^K)$. Since $\cD = \cD_{\cS,\cT} \times \cD_{\cS^c,\cT^c}$, this implies that
\begin{align*}
    &C([L]) - R([K]) = C(\cT) -R(\cS) + C(\cT^c) - R(\cS^c)\\
    &= I(Y_\cT;\Yh_\cT \cond X_\cS) + I(Y_{\cT^c};\Yh_{\cT^c} \cond X_{\cS^c})\\
    &=I(Y_\cT;\Yh_\cT) + I(Y_{\cT^c};\Yh_{\cT^c})- I(X_\cS;\Yh_\cT)  - I(X_{\cS^c};\Yh_{\cT^c}).\\[-2em]
\end{align*}
Note $I(Y_1^L;\Yh_1^L\cond X_1^K) = I(Y_\cT;\Yh_\cT\cond X_1^K) + I(Y_{\cT^c};\Yh_{\cT^c}\cond X_1^K)$ $=I(Y_\cT;\Yh_\cT) + I(Y_{\cT^c};\Yh_{\cT^c})- I(X_\cS;\Yh_\cT)-I(X_{\cS^c};\Yh_\cT \cond X_\cS)   - I(X_{\cS^c};\Yh_{\cT^c}) - I(X_\cS;\Yh_{\cT^c} \cond X_{\cS^c}).$
% \begin{align*}
%     &I(Y_1^L;\Yh_1^L\cond X_1^K) = I(Y_\cT;\Yh_\cT\cond X_1^K) + I(Y_{\cT^c};\Yh_{\cT^c}\cond X_1^K) \\
%     &=I(Y_\cT;\Yh_\cT) + I(Y_{\cT^c};\Yh_{\cT^c})- I(X_\cS;\Yh_\cT)-I(X_{\cS^c};\Yh_\cT \cond X_\cS)  \\
%     &\quad - I(X_{\cS^c};\Yh_{\cT^c}) - I(X_\cS;\Yh_{\cT^c} \cond X_{\cS^c}).
% \end{align*}
By equating the two equations, we have $I(X_\cS ; \Yh_{\cT^c} \cond X_{\cS^c}) = 0$ and $I(X_{\cS^c} ; \Yh_{\cT} \cond X_\cS) = 0$, and, thus, $(a)$ holds.

Now, we prove the other direction. By condition $(a)$, the independence of $X_\cS$ and $X_{\cS^c}$, and the fact that $\Yh_{\cT}$ and $\Yh_{\cT^c}$ are conditionally independent given $(X_1,\ldots,X_K)$, we have
\begin{equation} \label{eqn:independence}
    I(X_\cS,\Yh_\cT; X_{\cS^c},\Yh_{\cT^c}) = 0.
\end{equation}
Let $(R_1,\ldots,R_K,C_1,\ldots,C_L) \in \cD_{\cS,\cT}\times \cD_{\cS^c,\cT^c}$. By the same derivation as before, we have that $C([L]) - R([K]) = I(Y_1^L;\Yh_1^L\cond X_1^K)$. We need to further show that for every $\cA \subseteq [K]$, $\cB\subseteq [L]$, we have $C(\cB)-R(\cA) \leq I(Y_\cB;\Yh_\cB \cond X_\cA)$. Let $\cA = \cM \cup \cN$ and $\cB = \cP \cup \cQ$, where $\cM \subseteq \cS$, $\cN \subseteq \cS^c$, $\cP \subseteq \cT$ and $\cQ \subseteq \cT^c$. It follows that
\begin{align*}
    &C(\cB) - R(\cA) = C(\cP) - R(\cM) + C(\cQ) - R(\cN)\\
    &\leq I(Y_\cP;\Yh_\cP) - I(X_\cM;\Yh_\cP) + I(Y_\cQ;\Yh_\cQ ) - I(X_\cN;\Yh_\cQ)\\
    &= I(Y_{\cP \cup \cQ}; \Yh_{\cP\cup\cQ}) + I(\Yh_\cP;\Yh_\cQ)- I(X_\cM;\Yh_\cP)  - I(X_\cN;\Yh_\cQ)\\[-.5em]
    &\stackrel{(a)}{=} I(Y_{\cB}; \Yh_{\cB}) - I(X_{\cM\cup\cN}; \Yh_{\cP\cup\cQ})= I(Y_{\cB}; \Yh_{\cB} \cond X_\cA),\\[-2em]
\end{align*}
where $(a)$ holds by the fact any term in the chain rule expansion of~(\ref{eqn:independence}) should be zero. This implies that $(R_1,\ldots,R_K,C_1,\ldots,C_L) \in \cD$, and, thus, $\cD_{\cS,\cT}\times \cD_{\cS^c,\cT^c} \subseteq \cD$. Meanwhile, observation~(\ref{eqn:independence}) allows to express $\cD_{\cS,\cT}$ as
$
    \cD_{\cS,\cT} = \big\{(R_\cS,C_\cT): \forall \, \cA \subseteq \cS, \cB \subseteq \cT, 
     I(Y_\cB;\Yh_\cB \cond X_\cS, \Yh_{\cB^c}) - I(X_\cA;\Yh_{\cB^c}\cond X_{\cA^c})
     \leq C(\cB) - R(\cA) \leq I(Y_{\cB}; \Yh_{\cB} \cond X_{\cA})\big\}.$
% \begin{align*}
%     \cD_{\cS,\cT} &= \Big\{(R_\cS,C_\cT): \forall \, \cA \subseteq \cS, \cB \subseteq \cT, \\
%     &\qquad I(Y_\cB;\Yh_\cB \cond X_\cS, \Yh_{\cB^c}) - I(X_\cA;\Yh_{\cB^c}\cond X_{\cA^c}) \\
%     &\qquad \leq C(\cB) - R(\cA) \leq I(Y_{\cB}; \Yh_{\cB} \cond X_{\cA})\Big\}.
% \end{align*}
By the Fourier--Motzkin elimination method~\cite[Theorem 1.4]{Ziegler1995}, it can be shown that $\cD_{\cS,\cT}$ is the projection of $\cD$ onto the coordinates indexed by $(\cS,\cT)$. Similarly, $\cD_{\cS^c,\cT^c}$ is the projection of $\cD$ onto the coordinates indexed by $(\cS^c,\cT^c)$. Hence, $\cD\subseteq \cD_{\cS,\cT}\times \cD_{\cS^c,\cT^c}$. This completes the proof.

\vspace{-.5em}
\section{Proof of Lemma~\ref{lemma:degenerate-2}} \label{appx:degenerate-2}
Clearly, $(a)$ implies $(b)$ since
$
    \dim(\cD) = \dim(\cD_{\cS,\cT}) + \dim(\cD_{\cS^c,\cT^c})
    \leq \abs{\cS} + \abs{\cT} - 1 + \abs{\cS^c} + \abs{\cT^c} - 1 < K+L-1.$
% \begin{align*}
%     \dim(\cD) &= \dim(\cD_{\cS,\cT}) + \dim(\cD_{\cS^c,\cT^c})\\
%     &\leq \abs{\cS} + \abs{\cT} - 1 + \abs{\cS^c} + \abs{\cT^c} - 1 < K+L-1.
% \end{align*}
To show that $(b)$ implies $(a)$, we proceed by contradiction. Assume that $\cD \neq \cD_{\cS,\cT}\times \cD_{\cS^c,\cT^c}$ for each $\cS \subseteq [K]$ and $\cT \subseteq [L]$ such that $\cS \cup \cT \neq \emptyset$ and $\cS^c \cup \cT^c \neq \emptyset$. By Lemma~\ref{lemma:degenerate-1}, this means, for each allowed $(\cS,\cT)$, we either have $I(X_\cS ; \Yh_{\cT^c} \cond X_{\cS^c}) > 0$ or $I(X_{\cS^c} ; \Yh_{\cT} \cond X_\cS) > 0$. This implies
\begin{equation} \label{eqn:degenerate-strict}
    I(Y_\cT;\Yh_\cT \cond X_1^K) - I(X_\cS;\Yh_{\cT^c} \cond X_{\cS^c}) < I(Y_\cT;\Yh_\cT \cond X_\cS)
\end{equation}
for each $\cS\subseteq [K]$ and $\cT \subseteq [L]$ such that $\cS \cup \cT \neq \emptyset$ and $\cS^c \cup \cT^c \neq \emptyset$. Let $(R_1,\ldots,R_K,C_1,\ldots,C_L) \in \cD$. Then, $C_L$ is uniquely determined by $(R_1,\ldots,R_K,C_1,\ldots,C_{L-1})$. Let $\td{\cD}$ denote the restriction of $\cD$ to the first $K+L-1$ coordinates. Therefore, $\dim(\cD) = \dim(\td{\cD})$. For a given $\cS \subseteq [K]$ and $\cT \subseteq [L-1]$ such that $\cS\cup\cT \neq \emptyset$, let $\td{\cH}_{\cS,\cT}^{+}$ denote the hyperplane in $\mathbb{R}^{K+L-1}$ such that $C(\cT) -R(\cS) = I(Y_\cT;\Yh_\cT \cond X_\cS)$, and let $\td{\cH}_{\cS,\cT}^{-}$ denote the hyperplane in $\mathbb{R}^{K+L-1}$ such that
$
C(\cT) -R(\cS) = I(Y_\cT;\Yh_\cT \cond X_1^K) - I(X_\cS;\Yh_{[L-1]\setminus 
\cT} \cond X_{\cS^c}).
$
Let $\td{\cP}_{\cS,\cT}$ be the polyhedron bounded by $\td{\cH}_{\cS,\cT}^{+}$ and $\td{\cH}_{\cS,\cT}^{-}$, i.e., $\td{\cP}_{\cS,\cT}$ is defined by 
    $I(Y_\cT;\Yh_\cT \cond X_1^K) - I(X_\cS;\Yh_{[L-1]\setminus 
\cT} \cond X_{\cS^c}) 
    \leq C(\cT) -R(\cS) \leq I(Y_\cT;\Yh_\cT \cond X_\cS).$
% \begin{align*}
%     &I(Y_\cT;\Yh_\cT \cond X_1^K) - I(X_\cS;\Yh_{[L-1]\setminus 
% \cT} \cond X_{\cS^c}) \\
%     &\leq C(\cT) -R(\cS) \leq I(Y_\cT;\Yh_\cT \cond X_\cS).
% \end{align*}
Following~(\ref{eqn:degenerate-strict}), we have $\dim(\td{\cP}_{\cS,\cT}) = K+L-1$. Let
$
    \textstyle{\cC = \Big(\bigcap_{i \in [K]}\td{\cP}_{\{i\},\emptyset} \Big) \bigcap \Big(\bigcap_{j \in [L-1]}\td{\cP}_{\emptyset,\{j\}} \Big)}
$
denote the convex polytope bounded by constant hyperplanes. Following~(\ref{eqn:degenerate-strict}), we also have $\dim(\cC) = K+L-1$. By the Fourier--Motzkin elimination method~\cite[Theorem 1.4]{Ziegler1995}, it holds that
$
\textstyle{\td{\cD} = \bigcap_{\substack{\cS\subseteq [K], \, \cT\subseteq [L-1]: \cS\cup\cT \neq \emptyset}} \td{\cP}_{\cS,\cT}.}
$
Starting from $\cC$, we add each $\td{\cP}_{\cS,\cT}$, one by one, in order to construct $\td{\cD}$. We want to show that, at each step, the dimension of the resulting convex polytope remains $K+L-1$, just as the dimension of $\cC$. Let $\td{\cP}$ be the convex polytope at some step of this procedure, and let $\cS \subseteq [K]$ and $\cT \subseteq [L-1]$ be such that $\cS \cup \cT \neq \emptyset$. Define
$
    x_1 \in \td{\cP} \cap \td{\cH}_{\cS,\cT}^{+} \supseteq \td{\cD} \cap \td{\cH}_{\cS,\cT}^{+} \neq \emptyset,
    x_2 \in \td{\cP} \cap \td{\cH}_{\cS,\cT}^{-} \supseteq \td{\cD} \cap \td{\cH}_{\cS,\cT}^{-} \neq \emptyset,
$
and let $x = \frac{1}{2} (x_1+x_2)$. Thus, by~(\ref{eqn:degenerate-strict}), there exists an $\eps$ such that the ball $B_\eps(x)$ with center $x$ and radius $\eps$ is contained in $\td{\cP}_{\cS,\cT}$. But since $\td{\cP}$ is convex, then $x\in \td{\cP}$, and thus, $\dim(\td{\cP} \cap \td{\cP}_{\cS,\cT}) \geq \dim(\td{\cP} \cap B_\eps(x)) = K+L-1$. This implies that $\dim(\cD) = K+L-1$, which is a contradiction. 
% This completes the proof.
\end{appendices}

% \section*{Acknowledgement}
% The authors would like to express their deep gratitude to Prof. Young-Han Kim for numerous discussions around the problem of coding for cloud radio access networks.

\bibliography{bibliography}
\bibliographystyle{IEEEtran}

\end{document}